\documentclass[twocolumn,superscriptaddress,amsmath,amssymb,aps,pra]{revtex4-2}
\usepackage{graphicx}%
\usepackage{multirow}%
\usepackage{amsmath,amssymb,amsfonts}%
\usepackage{amsthm}%
\usepackage{mathrsfs}%
\usepackage{xcolor}%
\usepackage{textcomp}%
\usepackage{booktabs}%
\usepackage{algorithm}%
\usepackage{algorithmicx}%
\usepackage{algpseudocode}%
\usepackage{listings}%
\usepackage{braket}
\usepackage{float}
\usepackage{multirow}
\usepackage{hyperref}
\usepackage{microtype}
\usepackage{bm}
\newcommand{\Ba}{$^{138}$Ba$^+$~}
\newcommand{\DQC}{Duke Quantum Center, Duke University, Durham, NC 27708}
\newcommand{\IonQ}{IonQ, Inc., Boston, MA  02135}

\begin{document}

\title{Tripartite entanglement of remote atomic qubits}

\author{Isabella Goetting}
\email{isabella.goetting@duke.edu}
\affiliation{\DQC}
\author{Ashish Kalakuntla}
\affiliation{\DQC}
\author{Mikhail Shalaev}
\affiliation{\DQC}
\author{Harriet Bufan Shi}
\affiliation{\DQC}
\author{Ana Ferrari}
\affiliation{\DQC}
\author{Sagnik Saha}
\affiliation{\DQC}
\affiliation{\IonQ}
\author{George Toh}
\affiliation{\DQC}
\affiliation{\IonQ}
\author{Saki Male}
\affiliation{\DQC}
\author{Christopher Monroe}
\email{c.monroe@duke.edu}
\affiliation{\DQC}
\affiliation{\IonQ}

\begin{abstract} 
Distributed entanglement across multi-node quantum networks is essential for a wide range of quantum technologies, including modular quantum computers \cite{Monroe2014, 2007distqc}, 
%the quantum internet \cite{Kimble2008,2018quantuminternet}, 
distributed sensing and metrology \cite{2018quantumsensor,2020DistSensing, 2024SensingGuha, Komar2014}, and multi-party secure communication protocols \cite{Ekert1991, 2004MultipartyQSS, 1999quantunsecretsharing}.
Such large-scale quantum networks will require photonic interconnects to generate and sustain entangled states across localized nodes \cite{quantumrepeater1998}. 
Previously, three-node distributed Greenberger–Horne–Zeilinger (GHZ) states have been generated between solid-state qubits \cite{pompili2021} and atomic ensembles \cite{JingPan2019}, but not yet in the platform of individual atomic qubits, which can be replicated, detected, and individually controlled with high fidelity.
Here we report the first fully-distributed GHZ state of qubits across a three-node quantum network of single atomic memories, using photonic interconnects. We achieve a bounded fidelity of $0.841(17) \leq \mathcal{F} \leq 0.881(17)$ at an entanglement generation rate of 0.095(5)/sec and measure a clear violation of Mermin's inequality \cite{Mermin1990} while closing the detection loophole for the first time in a fully-distributed multipartite entangled state.

\end{abstract}

\maketitle

A scalable quantum computer architecture will ultimately require photonic interconnections between nodes of quantum memories \cite{Awschalom2021}.
Photonic interconnects provide a modular and reconfigurable quantum network that likely cannot be accomplished with just quantum memories and local gate operations. Such an architecture requires high-efficiency and high-fidelity optical interfaces with support for complex entanglement schemes within the nodes.

Remote entanglement of two qubit nodes has been accomplished in many optically-active qubit platforms such as trapped ions \cite{Moehring2007}, color centers in diamond \cite{Bernien2013}, quantum dots \cite{Delteil2016}, and neutral atoms \cite{Nolleke2013, Van_Leent_atom_ent_2022}. However, trapped ion modules have hosted the fastest photonic entanglement rates and highest fidelities \cite{Stephenson2020, OReilly2024, Saha2025, Drmota2023}, with unsurpassed performance in terms of quantum memory \cite{Zeemancoherence2016, cohertime2021}, qubit detection \cite{SPAM_Sotirova2024}, single-photon emission \cite{Crocker19}, and local quantum gate operations \cite{highFgates2025, highF1QG2025}.

Three-node distributed entanglement has been demonstrated in ensemble-based static memories, with a Greenberger-Horne-Zeilinger (GHZ) state fidelity of $0.71$ via interference of three photons at a rate of less than $0.002~\sec^{-1}$ \cite{JingPan2019}. 
Color centers in diamond have hosted a GHZ state fidelity of $0.54$ at a rate of $0.01~\sec^{-1}$, using a combination of photon-mediated Bell state entanglement in two separate nodes and local gates in a third, central node \cite{pompili2021}. 
Additionally, photonic platforms have demonstrated multipartite entanglement and violations of Mermin's inequality, closing the locality and freedom-of-choice loopholes \cite{Erven2014}.
While multipartite entangled states have been shown across two trapped ion nodes using a remotely generated Bell pair and local gates \cite{Main2025}, fully-distributed multipartite entanglement has not yet been achieved in this platform.

Here, we create a multi-node quantum network composed of three spatially-separated modules, each containing a single trapped ion qubit entangled with an emitted photonic qubit. We interfere the photons in a GHZ-state generator \cite{GHZPan1998} to herald the atomic qubits into a maximally-entangled GHZ state, 
\begin{equation} \label{eq:GHZ}
\ket{\Psi_{\text{GHZ}}^{\pm}} = \frac{\ket{\downarrow \downarrow \downarrow} \pm e^{i \Phi}\ket{\uparrow \uparrow \uparrow}}{\sqrt{2}}.
\end{equation}
We show the highest fidelity and the fastest rate of remote tripartite entanglement between three memory qubits. % using photonic interconnects.
The results are not post-selected and do not require any two-qubit gates, enabling an event-ready demonstration of a multi-node quantum network. 

A violation of Mermin's extension \cite{MerminIneq1990} of Bell's inequality has not previously been demonstrated across three distributed memories, in any platform.
In contrast to two-particle Bell tests, multipartite correlations can exhibit non-locality deterministically and without statistical averaging, allowing single trials to refute local realism. 
Pure photonic platforms have historically required additional assumptions, leaving open the possibility of loopholes such as the ``fair sampling" hypothesis associated with imperfect qubit detection \cite{Erven2014}.
Trapped atomic ions are well-suited to close the detection loophole because of their near-perfect detection efficiency \cite{2003SimonIrvine, Rowe2001}. Here we demonstrate a strong violation of Mermin's inequality and close the detection loophole for the first time in a fully-distributed system. 
\begin{figure}[h!]
  \includegraphics[width=0.5\textwidth]{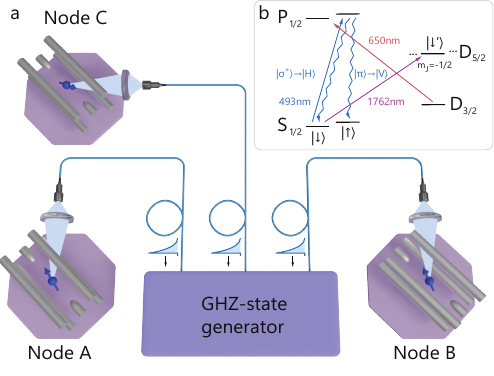}
  \caption{\small \textbf{Three-node network and atomic energy levels.}  (a) Schematic of three ion trap nodes, each separated by about $2$ m, and photonic optical fibers ($3$ m long) leading to the GHZ generator. Each ion is confined in a four-rod Paul trap. (b) Reduced energy level diagram of \Ba system. Fast excitation at 493 nm prepares the ion in the $P_{1/2}$ level and subsequent spontaneous emission ($73\%$ branching ratio) creates ion-photon entanglement between the Zeeman atomic qubit $\ket{\downarrow}_i/\ket{\uparrow}_i$ and the polarization state of the photon $\ket{H}_i/\ket{V}_i$. The qubit is measured by first shelving the state to the $D_{5/2}$ manifold $\ket{\downarrow}_i \rightarrow \ket{\downarrow^\prime}_i$ and then collecting fluorescence under illumination from both $493$ nm and $650$ nm light.}
  \label{fig:Schematic}
\end{figure}
\section*{Entanglement across the three-node network}

The multi-node quantum network is composed of three spatially-separated trapped ion modules denoted by $A, B, \text{and }C$, each containing a single \Ba ion qubit, as depicted in Fig.\ref{fig:Schematic}a and detailed in Methods. 
A magnetic field of $B = 4.2446(2)$ G at each node lifts the degeneracy of the ground-state, defining the atomic Zeeman qubit levels $\ket{\downarrow}_i \equiv \ket{^2S_{1/2}, m_J = -1/2}$ and $\ket{\uparrow}_i \equiv \ket{^2S_{1/2}, m_J = +1/2}$, split by a frequency of $\omega_0= 2\pi\times 11.8964(5)$~MHz that is matched to within 1 kHz for each qubit (see Fig. \ref{fig:Schematic}b).

The entanglement sequence begins with the initialization of each trapped ion qubit to the $\ket{\downarrow}_i$ state followed by a fast ($\sim 3$ ps) and simultaneous laser excitation of each trapped ion to their $\ket{^2P_{1/2}, m_J = +1/2}$ excited state (lifetime $\tau\sim 8$ ns), with excitation probability $> 0.8$. Subsequent spontaneous emission at 493 nm back to the ground state ideally creates an entangled state between trapped ion qubit $i$ and its emitted photon \cite{OReilly2024}
\begin{equation}
    \ket{\Psi}_{i} = \frac{\ket{H}_i\ket{\downarrow}_i + e^{i \delta k_i x_i} \ket{V}_i\ket{\uparrow}_i}{\sqrt{2}}, \label{eq:IP}
\end{equation}
given that the photon is collected.
In the above state, $\ket{H}_i$ and $\ket{V}_i$ refer to the horizontal/vertical polarization states of the photonic qubit, $\delta k_i\sim 0.4$ m$^{-1}$ is the difference in the emission wavenumbers between the $H$ and $V$ photons, and $x_i$ is the photonic path length.
High numerical-aperture (NA) lenses are used to collect each of the three single photons into single-mode optical fibers directed perpendicularly to the magnetic field at each node, so that $\sigma^+/\pi$-emission is projected to a $H/V$ photon \cite{Stephenson2020,OReilly2024}. The average fidelity of the three ion-photon states is 0.983(1), limited mainly by polarization mixing. 
\begin{figure}
    \centering\includegraphics[width=\linewidth]{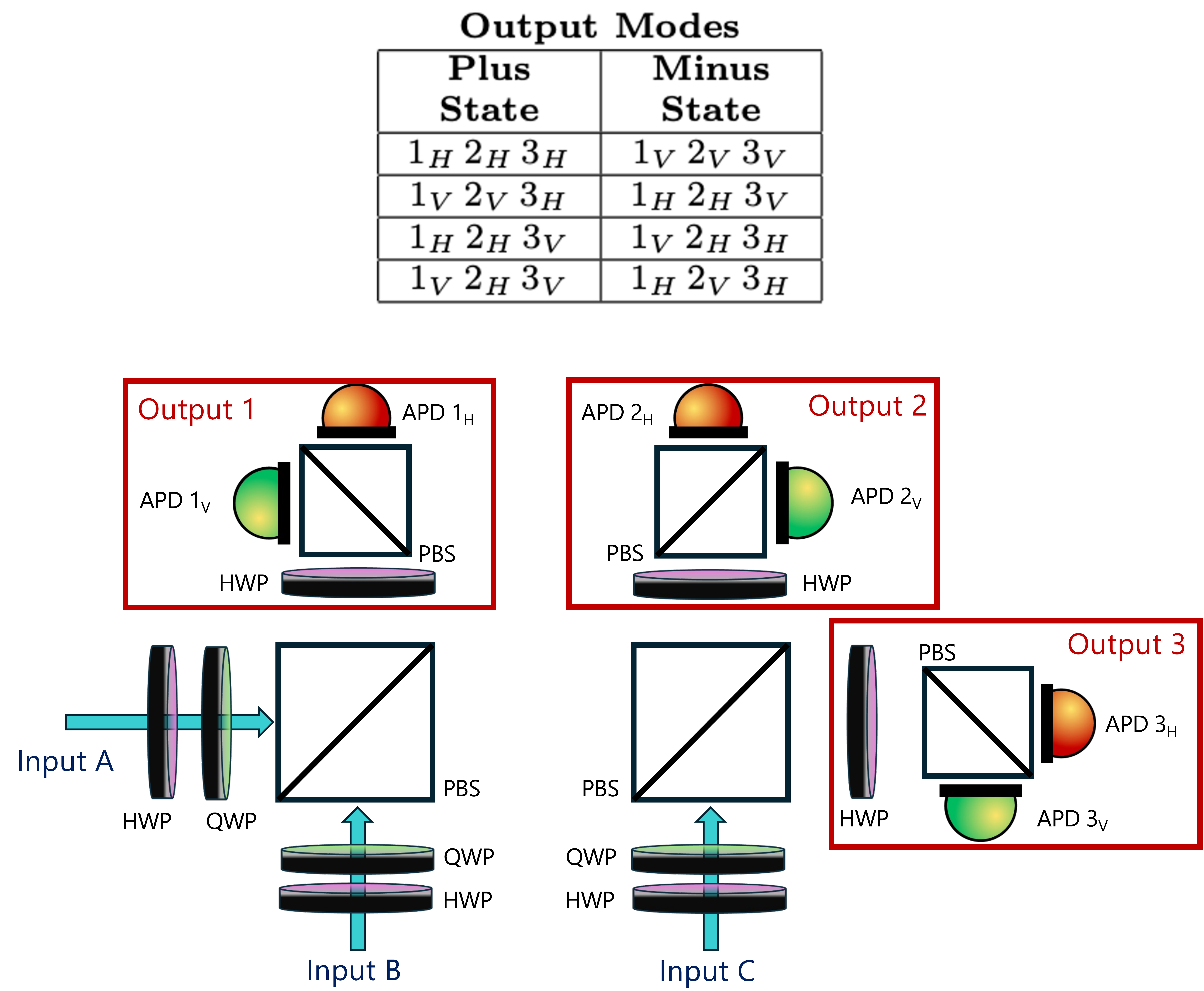}
    \caption{\small \textbf{Details of GHZ-state generator.}  The input modes A, B, and C carry the input photons from the respective trapped ion qubits. The photons travel through half-waveplate (HWP/purple) and quarter-waveplate (QWP/green) stacks, which match the photonic qubits to the same horizontal/vertical (H/V) basis. The photons interfere pairwise at input polarizing beam splitters (PBS). 
    At the output modes labeled $1$, $2$, and $3$, half-wave plates rotate the photons into the $H\pm V$ bases and erase ``which-path" information. Output PBS cubes separate H and V photons and direct them to single photon avalanche photodiodes (APD) labeled by their respective mode and polarization. 
    The table on top indicates the eight triple-coincidence detection patterns that herald the plus or minus GHZ states.
    }
    \label{fig:GHZsetup}
\end{figure}

After generating ion-photon states among all three nodes according to Eq. \ref{eq:IP}, the single photons travel 3~m in their respective fibers to the GHZ-state generator \cite{GHZPan1998} depicted in Fig. \ref{fig:GHZsetup}. Here, wave plates first rotate the polarization of the incoming photons to align along the same $H/V$ basis. Then the photons are routed to three output modes through the two central polarizing beam splitters (PBS).
After this partitioning, a half-wave plate (HWP) at each of the three output modes rotates the photons into the $\ket{H}\pm \ket{V}$ bases, erasing the ``which-path" information of each photon. These photons must be indistinguishable in temporal and spatial mode, frequency, and polarization for optimal Hong-Ou-Mandel  \cite{HOM} interference and high-fidelity entanglement.

The three output photons are finally detected with six avalanche photodiodes (APD); we keep only the eight cases in which a single photon (H or V) is detected in each of the three output modes (see Fig. \ref{fig:GHZsetup}). 
This projects the atomic qubits into one of the GHZ states of Eq.~\eqref{eq:GHZ}, with the sign of the heralded state determined by the three-photon detection pattern.
The table in Fig. \ref{fig:GHZsetup} outlines all eight successful three-fold coincidence detection patterns. The GHZ phase in Eq. \ref{eq:GHZ} is ${\Phi = \delta k_Ax_A + \delta k_Bx_B + \delta k_Cx_C - \omega_0 (t_{1} + t_{2} + t_{3})}$. Here, $t_j$ is the recorded time of the photon click in output mode $j$ with reference to the first photon detected at ${t_j \equiv 0}$. 
Once a successful three-photon coincidence is detected within a 50 ns window, we proceed to state analysis as described below. Otherwise, entanglement attempts are continuously repeated with periodic breaks for Doppler cooling (see Methods). 

\section*{State Analysis and Entanglement Characterization}

To measure the state populations of each ion qubit, a narrow-linewidth 1762~nm laser drives a $\pi$-pulse on the shelving transition of each ion $\ket{\downarrow}$ to $\ket{\downarrow'}\equiv \ket{^2D_{5/2}, m_J = -1/2}$. Then, each ion is exposed to 493~nm and 650~nm light for 1.5~ms and the resulting fluorescence is collected (see Fig. \ref{fig:Schematic}b). Ion $i$ fluoresces when in state $S_i = \ket{\uparrow}$ and remains dark when in state $S_i = \ket{\downarrow}$, resulting in a state detection fidelity of $> 99.7(2) \%$ per node \cite{Saha2025}. From these three independent measurements, we obtain the eight state populations $\Pi_{S_A S_B S_C}$. 

We perform partial state tomography to bound the fidelity of the distributed GHZ state with respect to the ideal target state $\ket{\Psi_{\text{GHZ}}}$. 
In addition to the state populations, we measure the parity contrast 
$\mathcal{C}$ in a rotated basis: a $\pi/2$ pulse is applied to each 
qubit before measurement, and the qubit parity is computed. 
When the phase $\phi$ of the $\pi/2$ pulse on any one qubit is scanned, the parity oscillates as $\pm\mathcal{C}\cos(\phi + \Phi + \Phi_0)$. The sign follows from the nominal GHZ state created in Eq. \ref{eq:GHZ} and the constant phase offset $\Phi_0$ is accumulated during the time between photon detection and the subsequent coherent qubit analysis operations (see Appendix \ref{phasetrackingsupp}).

The GHZ state fidelity $\mathcal{F}$ is bounded by
\begin{equation}
\frac{\mathcal{P}+\mathcal{C}}{2}-\mathcal{Q}\le \mathcal{F} \le \frac{\mathcal{P}+\mathcal{C}}{2}+\mathcal{Q}, 
\end{equation}
as shown in Appendix \ref{fidboundsupp}.
Here, $\mathcal{P}=\Pi_{\downarrow\downarrow\downarrow} + \Pi_{\uparrow\uparrow\uparrow}$ is the sum of nominal state populations and $\mathcal{Q}=\sqrt{\Pi_{\downarrow\uparrow\uparrow}\Pi_{\uparrow\downarrow\downarrow}}
+\sqrt{\Pi_{\downarrow\downarrow\uparrow}\Pi_{\uparrow\uparrow\downarrow}}
+\sqrt{\Pi_{\downarrow\uparrow\downarrow}\Pi_{\uparrow\downarrow\uparrow}}$ represents the error populations.
The measured populations and parity fringes (see Appendix \ref{paritysupp}) are displayed in Fig. \ref{fig:pops}, 
resulting in a GHZ state fidelity $0.841(17) \leq \mathcal{F} \leq 0.881(17)$. 
We note that the fidelity bounds can be made tighter
by scanning the phase of all three qubits in the rotated-basis parity measurement and extracting the contrast of the oscillation with $3\phi$ \cite{Sackett2000}.

\begin{figure}[h!]
  \centering
  \begin{minipage}[]{0.5\textwidth} \centering %\includegraphics[width=\textwidth]{pops_inset_concat.pdf}
  \includegraphics[width=\textwidth]{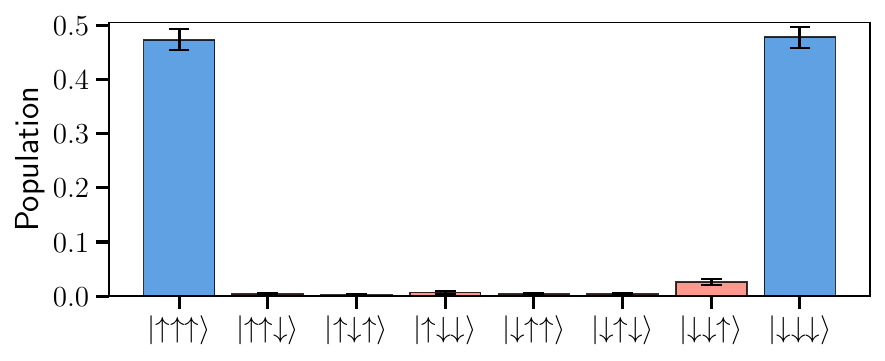}
  \end{minipage}
    \begin{minipage}[]{0.5\textwidth} \centering 
  \includegraphics[width=\textwidth]{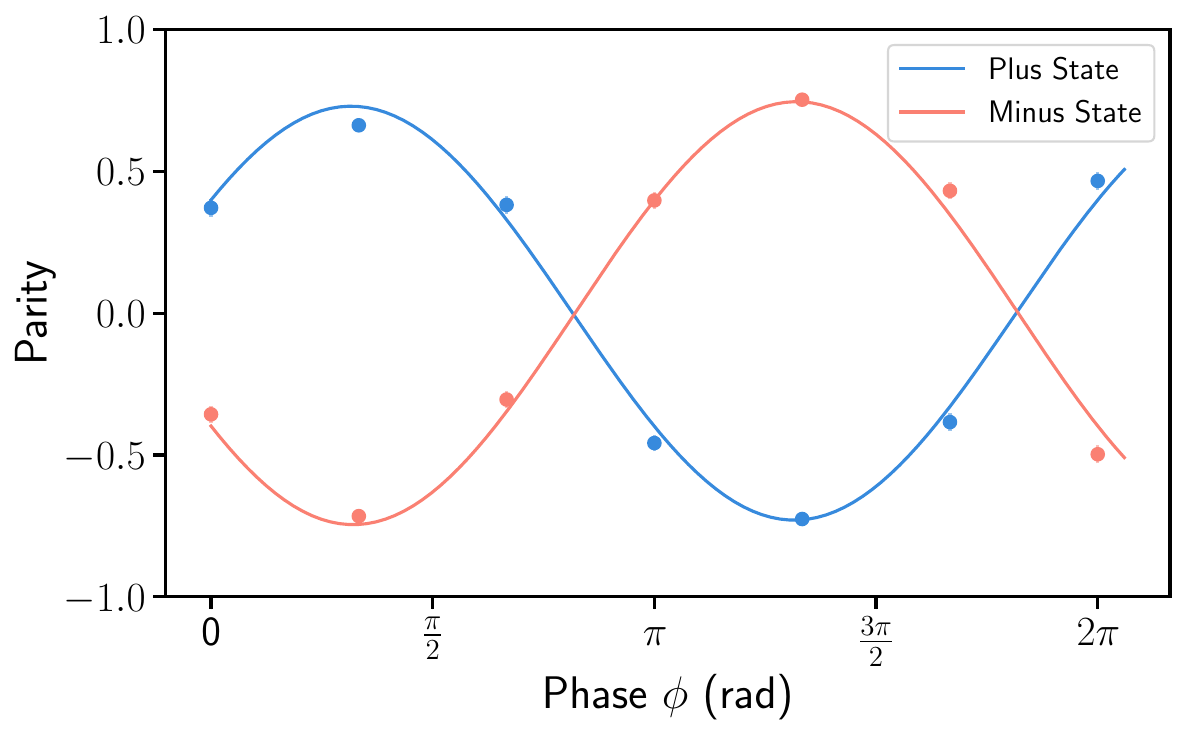}
  \end{minipage}
  \caption{ \small \textbf{GHZ state characterization.} (TOP) Measured populations $\Pi_{S_A S_B S_C}$ averaged over both plus and minus GHZ states, giving nominal populations (blue) with $\mathcal{P}=0.955(8)$ and error population (red) with $\mathcal{Q}=0.020(4)$ (statistical errors). Population data are collected from 687 successes out of 4,110,118,427 attempts, occupying about 2.7 hours of wall clock time. (BOTTOM) Rotated-basis parity of plus (blue) and minus (red) GHZ states, as the phase $\phi$ of one of the qubit rotations is scanned, resulting in average contrast $\mathcal{C}=0.77(2)$ with a fitted phase $\Phi + \Phi_0 = -1.00(5)$ rad.  
  Combining populations and parity contrast, we can bound the GHZ fidelity by $0.841(17) \leq \mathcal{F} \leq 0.881(17)$. 
  Parity fringe data are collected from 3110 successes out of 19,153,881,819 attempts, occupying about 12 hours of wall clock time.
 }
  \label{fig:pops}
\end{figure}

\begin{table}[h!]
\caption{Estimated error budget for the GHZ state, with the expected net infidelity consistent with observation. 
}
\begin{tabular}{|c|r|}
\hline
\textbf{Error source}&\textbf{Infidelity} \\ \hline\hline
Polarization mixing& 0.037(3)  \\ \hline
Spatial mode mismatch& \textless{}$0.03(1)$\phantom{7}\\ \hline
Recoil error & 0.03(2)\phantom{7} \\ \hline
SPAM& 0.015\phantom{(3)} \\ \hline Temporal overlap & \textless{}0.006\phantom{(3)}\\ \hline
Phase tracking & \textless{}0.005\phantom{(3)}\\ \hline
Wave plate calibration & \textless{}0.005\phantom{(3)}\\ \hline
Magnetic field mismatch & \textless{}0.001\phantom{(3)}\\ \hline
\textbf{Total error}    & 0.12(2)\phantom{7}
\\ \hline
\end{tabular}
\label{taberror}
\end{table}
As shown in Table \ref{taberror}, we attribute the main sources of error in the GHZ-state fidelity to photonic spatial mode mismatch, polarization mixing, and recoil decoherence, discussed below in turn. The product of the pairwise visibilities of the three photonic inputs is $0.941$, according to auxiliary interferometric measurements. The resulting residual photon distinguishability would contribute a maximum GHZ-state infidelity of $0.03(1)$ (see Appendix \ref{indistinguishabilitysupp} for derivation). 
Ion-photon coherence error is mainly due to polarization mixing, with a small contribution from state preparation and measurement (SPAM) errors, contributing an infidelity of $0.052(3)$. 
The other dominant error arises from the residual entanglement with motion due to recoil over the 50~ns photon heralding window \cite{Saha2025,Yu2026}. We estimate a parity contrast degradation of $0.06$ from this effect, resulting in an additional infidelity of $0.03(2)$ on the GHZ state (see Figure  \ref{fig:ionion} in Appendix \ref{ionionsupp}). 
A detailed error budget is included in the Appendix \ref{errorbudgetsupp}.

The overall tripartite entanglement generation rate is $r_{\text{ent}} = \frac{1}{4} R D p_A p_B p_C$. Here, $R$ is the entanglement attempt rate, $D$ is the duty cycle of attempted photon generation to allow for intermittent laser cooling of the ion, and the factor of $1/4$ comes from the fraction of GHZ states we can herald using linear optics. 
The individual end-to-end photon collection efficiencies $p_i$ account for non-unit excitation probability and all losses in the photonic channel from light collection to detector efficiency (Table \ref{photoneff}). 
Given $R=1$~MHz and $D=0.625$, 
we observe a net tripartite entanglement generation rate of $r_\text{ent}=0.095(8) \sec^{-1}$ (see Appendix \ref{GHZratesupp}).

We note that increasing the duty cycle does not necessarily improve the entanglement rate, as accumulated recoil heating from photon generation can degrade the photon collection efficiency into the fiber and increase the likelihood of ion de-crystallization. 
However, these de-crystallization events can be eliminated with a duty cycle approaching $100\%$ by employing sympathetic cooling with a second ion in each node \cite{sympathetic,OReilly2024} or using a faster cooling method such as dark-resonance cooling \cite{Allcock_2016, EIT2016}. This improvement could recover the uninterrupted entanglement rate of $0.25 \sec^{-1}$. 

\begin{table}[h!]
\caption{Breakdown of photon collection efficiency factors for the three nodes. The measured net efficiency is $36\%$ lower than expected, which we attribute to ion de-crystallization events and fiber coupling drift between calibrations.
}
\begin{tabular}{|c|c|c|c|}
\hline
\textbf{\begin{tabular}[c]{@{}c@{}}Photon collection \\ efficiency factors\end{tabular}}
 & \textbf{\begin{tabular}[c]{@{}c@{}}Node\\ A\end{tabular}} & \textbf{\begin{tabular}[c]{@{}c@{}}Node\\ B\end{tabular}} & \textbf{\begin{tabular}[c]{@{}c@{}}Node\\ C\end{tabular}} 
 \\ \hline\hline
Optical pumping& 0.92 & 0.92 & 0.94  \\ \hline
Excitation probability & 0.84 & 0.90 & 0.75  \\ \hline
Branching ratio & 0.73 & 0.73 & 0.73 \\ \hline
Collection solid-angle& 0.10 & 0.10 & 0.20  \\ \hline
Trap transmission & 0.78 & 0.78 & 0.97 \\ \hline
Fiber coupling & 0.35 & 0.26 & 0.23 \\ \hline
Detector efficiency & 0.68 & 0.66 & 0.70 \\ \hline
Optical transmission & 0.90 & 0.90 & 0.90 \\ \hline
\textbf{Net efficiency $p_i$} &  $0.0094$ & $0.0074$ & $0.0145$ \\ \hline 
\multicolumn{1}{|c|}{\textbf{Expected $p_A p_B p_C$}} & \multicolumn{3}{c|}{$1.01 \times 10^{-6}$} \\ \hline
\multicolumn{1}{|c|}{\textbf{Measured $p_A p_B p_C$}} & \multicolumn{3}{c|}{$0.636(12) \times 10^{-6}$} \\ \hline
\end{tabular}
\label{photoneff}
\end{table}

\section*{Violation of Mermin's inequality}
Bell inequalities certify that quantum entanglement produces correlations no classical, locally causal theory can reproduce. For two particles, the maximum quantum violation exceeds the classical bound by a factor of $\sqrt{2}$, revealed only through statistical accumulation over many runs. Mermin showed that the violation grows exponentially with the number of particles $n$ and, for a perfect GHZ state, is revealed by single trials rather than a statistical argument \cite{MerminIneq1990}. The Mermin inequality thus provides a natural and stringent benchmark for multipartite entanglement.

We violate Mermin's inequality for the first time in distributed memories. Previous distributed photonic demonstrations needed to invoke the fair-sampling assumption \cite{Erven2014}, leaving the so-called ``detection loophole" open. The fair-sampling assumption presumes that the subset of detected measurement outcomes is a representative sample of all outcomes, including those that went undetected \cite{fairsampling}. Because of the near-unity state detection efficiency, we also close the detection loophole. 

The Mermin inequality for three particles is
\begin{equation}
M  = |\braket{YYX} + \braket{ YXY} + \braket{XYY} - \braket{XXX}| \leq 2,    \label{eq:Mermin}
\end{equation}
where the operators $X,Y$ are the Pauli matrices $\sigma_x, \sigma_y$.
The above bound arises for any local hidden variable theory, while quantum mechanics is bounded by $M \leq 4$.

\begin{figure}[h!]
  \centering
  \includegraphics[width=\linewidth]{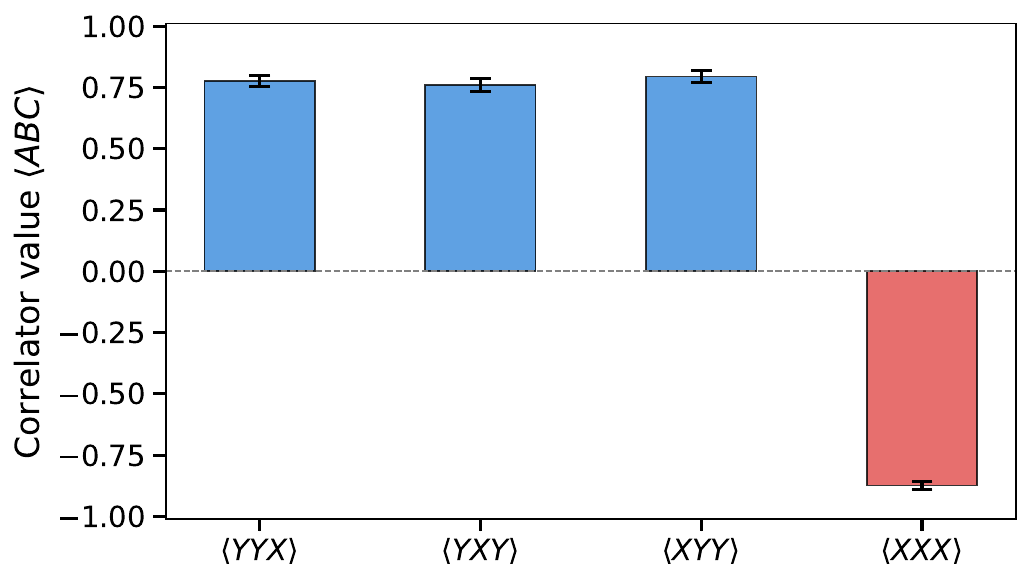}
  \caption{\textbf{Measured correlators of the Mermin parameter.} The measurements represent 735 events out of 4,883,403,222 attempts, taking 2.5 hours of wall clock time. We measure an average Mermin parameter of $M=3.203(45)$, with a purely statistical error bar.}
  \label{fig:mermin}
\end{figure}

Once a three-photon event is heralded, we measure any of the four correlators in Eq. \ref{eq:Mermin} by first shelving each qubit state $\ket{\downarrow}_i$ to the $\ket{\downarrow^\prime}_i$ state to create optical qubits as above. Before fluorescence detection, we apply 1762~nm $\pi/2$ pulses on each qubit, driving the $\ket{\uparrow}_i\leftrightarrow\ket{\downarrow^\prime}_i$ qubit transitions. 
The phases of each optical $\pi/2$ pulse determine the basis (X or Y) of the subsequent fluorescence measurement for each qubit.
When a plus (minus) GHZ state has been heralded, we feed-forward a factor of $+1$ ($-1$) to each measured correlator.
The measurements are displayed in Fig. \ref{fig:mermin}, resulting in a Mermin parameter of $3.203(45)$, violating Mermin's inequality by 27 standard deviations. 

\section*{Outlook}
\vspace{-6pt}
The rate of tripartite entanglement demonstrated in this work scales with the individual photon collection efficiency as $p_i^3$, and is therefore a natural approach for systems with $p_i$ approaching unity, such as those involving high-cooperativity cavities \cite{Nolleke2013}. Nevertheless, even though $p_i \sim 1\%$, the work presented here is the fastest rate of remote tripartite entanglement using photonic interconnects, and at the highest fidelity.
Several straightforward modifications could significantly enhance the rate and fidelity of entanglement generation. In addition to faster or continuous cooling methods discussed above, integrated optical elements such as in-fiber beamsplitters and PBS elements are expected to improve performance. 

%\red{Could argue that we never create a 3-photon interference...}
Two-photon protocols can create a GHZ state, by creating Bell pairs between nodes $A/C$ and $B/C$, followed by local operations at central node $C$ \cite{pompili2021}. Given this rate scales as $p_i^2$ instead of $p_i^3$, it can enable faster remote GHZ-state generation and lays the infrastructure for a quantum repeater \cite{quantumrepeater1998}. However, creating a GHZ state via three photons may still reduce overhead compared to making multiple Bell states.
Another variation is to use a three-port (fiber) beam splitter, such as a tritter \cite{Wstate1997,tritter}, which enables the creation of a  W-state, $\ket{W} = \ket{\downarrow \downarrow \uparrow} + \ket{\downarrow \uparrow \downarrow} + \ket{\uparrow \downarrow \downarrow}$, a tripartite entangled state belonging to a different entanglement class \cite{Kumar2023}. 

Finally, we note that the remote tripartite entangled system reported here can facilitate optimal protocols for distributed sensing \cite{2019Gorshkov, 2021Alexey} as well as
novel multi-party communication and cryptography protocols. These include three-party conference key agreement \cite{confkey}, secure quantum secret sharing \cite{2004MultipartyQSS}, non-local gates for distributed quantum computing \cite{2007distqc}, and multiparty secure random number generation \cite{threepartyRNG}. 
These and other applications are especially compelling in trapped ion systems, which enjoy high-fidelity local entanglement with neighboring trapped ion qubits through conventional Coulomb-based gates as well as near-perfect detection efficiency.
\vspace{-6pt}
\section*{Methods}\label{methods}
\vspace{-6pt}
\textbf{Three-node network.}
Each module of the three-node network consists of an ultra-high vacuum chamber containing a single \Ba ion in a four-rod Paul trap. Nodes $A, B,$ and $C$ are arranged roughly in a line with  $\sim 2$~m separation between neighboring chambers. Emitted single photons are collected into single-mode fibers using free-space high-numerical aperture (NA) lenses: nodes $A/B$ use NA=0.6 objectives ($10 \%$ collection solid angle), and node $C$ uses an in-vacuum NA=0.8 asphere ($20\%$ solid angle). The magnetic fields at each ion are calibrated to within 1~kHz of each other by doing narrow-linewidth spectroscopy on the transition $\ket{\downarrow'}$ to $\ket{\uparrow}$ and accordingly using a frequency-to-voltage conversion to adjust the voltage applied to the magnetic field compensation coils of each node. To minimize phase drift between runs, the 1762~nm pulse times remain the same and the magnetic fields are re-calibrated to the original values.

\textbf{Experimental sequence.}
The experimental sequence begins by Doppler cooling the ions for $150~ \mu$s with 493~nm light, followed by a series of entanglement attempts, each $1~\mu$s long. First, the ions are optically pumped to $\ket{\downarrow}$ using 493~nm and 650~nm light. Next, the ions are simultaneously excited with a 3~ps pulse of circularly-polarized 493~nm light to generate single photons with probability $> 80 \%$ (limited by laser power), followed by a 50~ns single-photon detection window. 
Because the 8~ns lifetime of the excited $^2P_{1/2}$ state is much longer than the 3~ps 493~nm pulse, the probability of generating a double excitation is $<10^{-3}$. The single photon spontaneously decays with $73\%$ probability to the $^2S_{1/2}$ manifold, resulting in $\pi$-emission to $\ket{\uparrow}$ or $\sigma^+$-emission to $\ket{\downarrow}$. When collected perpendicular to the magnetic field and into single-mode optical fiber, the $\sigma^+$ photon gets projected to the horizontal polarization mode (H) and $\pi$ to vertical (V) with equal probability, generating the entangled ion-photon state \ref{eq:IP}. 
The photons each travel 3~m in fiber to the GHZ-state generator, where the photons are interfered in free-space. 

\textbf{State Analysis.}
If a successful three-photon coincidence is detected as described above, we proceed to state analysis and detection. Otherwise, attempts are continuously repeated for up to 250-350 cycles at $1~\mu$s per cycle before a $150~\mu$s interruption for Doppler cooling, for a duty cycle of $62.5-70\%$. This Doppler cooling and entanglement attempt sequence is repeated 100 times ($<50$ ms) per shot, with a total of 16,000 shots averaged per parity point and 5,000 shots per population point. 

The time between a successful three-photon detection event and the start of the analysis pulse sequence is $3.5~\mu$s, ensuring negligible loss from qubit decoherence. To measure the populations, a 1762~nm $\pi$-pulse shelves $\ket{\downarrow}$ to $\ket{\downarrow'}\equiv \ket{^2D_{5/2}, m_J = -1/2}$ for approximately $6 \mu$s, followed by 1.5~ms of conventional fluorescence detection using 493~nm and the 650~nm and 614~nm repumpers. To measure parity, the same 1762 nm shelving pulse first transforms the ground-state qubit into an optical qubit. After a few $\mu$s, a subsequent 1762 nm $\pi/2$ pulse spanning $\ket{\uparrow}$ to $\ket{\downarrow'}$ is performed, where one ion receives an ``analysis" phase and the other two receive zero phase. The analysis phase is scanned over a full period to obtain the parity fringe. 
To measure the Mermin correlators, we phase shift the 1762 nm $\pi/2$ pulse by either 0 or $\pi/2$ for an $X$ or $Y$ measurement of each ion.
\vspace{-6pt}
\section*{Acknowledgments}
\vspace{-6pt}
This work is supported by the DOE Quantum Systems Accelerator (DE-FOA-0002253) and the NSF STAQ Program (PHY-1818914). A.K. is supported by the AFOSR National Defense Science and Engineering Graduate (NDSEG) Fellowship. After the completion of this manuscript, we learned of a recent experimental result on tripartite remote entanglement of single neutral atoms, in the group of G. Rempe (Max Planck Institute for Quantum Optics, Garching, Germany).

\bibliographystyle{apsrev4-1}
\bibliography{bib}% common bib file

@article{Wstate1997,
  title = {Realizable higher-dimensional two-particle entanglements via multiport beam splitters},
  author = {\ifmmode \dot{Z}\else \.{Z}\fi{}ukowski, Marek and Zeilinger, Anton and Horne, Michael A.},
  journal = {Phys. Rev. A},
  volume = {55},
  issue = {4},
  pages = {2564--2579},
  numpages = {0},
  year = {1997},
  month = {Apr},
  publisher = {American Physical Society},
  doi = {10.1103/PhysRevA.55.2564},
  url = {https://link.aps.org/doi/10.1103/PhysRevA.55.2564}
}

@article{pompili2021,
  author  = {Pompili, M. and Hermans, S. L. N. and Baier, S. and Beukers, H. K. C. and Humphreys, P. C. and Schouten, R. N. and Vermeulen, R. F. L. and Tiggelman, M. J. and Martins, L. dos Santos and Dirkse, B. and Wehner, S. and Hanson, R.},
  title   = {Realization of a multinode quantum network of remote solid-state qubits},
  journal = {Science},
  volume  = {372},
  number  = {6539},
  pages   = {259--264},
  year    = {2021},
  doi     = {10.1126/science.abg1919}
}

@article{JingPan2019,
  author  = {Jing, Bo and others},
  title   = {Entanglement of three quantum memories via interference of three single photons},
  journal = {Nat. Photonics},
  volume  = {13},
  pages   = {210--216},
  year    = {2019},
  doi     = {10.1038/s41566-018-0342-x}
}

@article{GHZPan1998,
   author = {Pan, J. W. and Zeilinger, A.},
   title = {Greenberger-Horne-Zeilinger-state analyzer},
   journal = {Phys. Rev. A},
   volume = {57},
   number = {3},
   pages = {2208},
   ISSN = {2469},
   DOI = {DOI 10.1103/PhysRevA.57.2208},
   url = {<Go to ISI>://WOS:000072441900093
https://journals.aps.org/pra/pdf/10.1103/PhysRevA.57.2208},
   year = {1998},
   type = {Journal Article}
}

@article{Main2025,
      title={Multipartite Mixed-Species Entanglement over a Quantum Network}, 
      author={D. Main and others},
      year={2025},
      journal={arXiv:2506.14334},
      url={https://arxiv.org/abs/2506.14334}, 
}

@article{Erven2014,
   author = {Erven, C. and others},
   title = {Experimental three-photon quantum nonlocality under strict locality conditions},
   journal = {Nature Photonics},
   volume = {8},
   number = {4},
   pages = {292–296},
   ISSN = {1749-4885},
   DOI = {10.1038/Nphoton.2014.50},
   url = {<Go to ISI>://WOS:000333800000009
https://www.nature.com/articles/nphoton.2014.50.pdf},
   year = {2014},
   type = {Journal Article}
}

@article{Ekert1991,
  title = {Quantum cryptography based on Bell's theorem},
  author = {Ekert, Artur K.},
  journal = {Phys. Rev. Lett.},
  volume = {67},
  issue = {6},
  pages = {661--663},
  numpages = {0},
  year = {1991},
  month = {Aug},
  publisher = {American Physical Society},
  doi = {10.1103/PhysRevLett.67.661},
  url = {https://link.aps.org/doi/10.1103/PhysRevLett.67.661}
}

@article{2007distqc,
  title = {Distributed quantum computation based on small quantum registers},
  author = {Jiang, Liang and Taylor, Jacob M. and S\o{}rensen, Anders S. and Lukin, Mikhail D.},
  journal = {Phys. Rev. A},
  volume = {76},
  issue = {6},
  pages = {062323},
  numpages = {22},
  year = {2007},
  month = {Dec},
  publisher = {American Physical Society},
  doi = {10.1103/PhysRevA.76.062323},
  url = {https://link.aps.org/doi/10.1103/PhysRevA.76.062323}
}

@article{1999quantunsecretsharing,
  title = {Quantum secret sharing},
  author = {Hillery, Mark and Bu\ifmmode \check{z}\else \v{z}\fi{}ek, Vladim\'{\i}r and Berthiaume, Andr\'e},
  journal = {Phys. Rev. A},
  volume = {59},
  issue = {3},
  pages = {1829--1834},
  numpages = {0},
  year = {1999},
  month = {Mar},
  publisher = {American Physical Society},
  doi = {10.1103/PhysRevA.59.1829},
  url = {https://link.aps.org/doi/10.1103/PhysRevA.59.1829}
}

@article{Komar2014,
   author = {Kómár, P. and Kessler, E. M. and Bishof, M. and Jiang, L. and Sorensen, A. S. and Ye, J. and Lukin, D.},
   title = {A quantum network of clocks},
   journal = {Nature Phys.},
   volume = {10},
   number = {8},
   pages = {582},
   ISSN = {1745-2473},
   DOI = {10.1038/Nphys3000},
   url = {<Go to ISI>://WOS:000340140300015
https://www.nature.com/articles/nphys3000.pdf},
   year = {2014},
   type = {Journal Article}
}

@article{2020DistSensing,
   author = {Guo, X. S. and Breum, C. R. and Borregaard, J. and Izumi, S. and Larsen, M. and Gehring, T. and Christandl, M. and Neergaard-Nielsen, J. S. and Andersen, U. L.},
   title = {Distributed quantum sensing in a continuous-variable entangled network},
   journal = {Nature Phys.},
   volume = {16},
   number = {3},
   pages = {281},
   ISSN = {1745-2473},
   DOI = {10.1038/s41567-019-0743-x},
   url = {<Go to ISI>://WOS:000507726300003
https://www.nature.com/articles/s41567-019-0743-x.pdf},
   year = {2020},
   type = {Journal Article}
}

@article{Monroe2014,
  title = {Large-scale modular quantum-computer architecture with atomic memory and photonic interconnects},
  author = {Monroe, C. and Raussendorf, R. and Ruthven, A. and Brown, K. R. and Maunz, P. and Duan, L.-M. and Kim, J.},
  journal = {Phys. Rev. A},
  volume = {89},
  issue = {2},
  pages = {022317},
  numpages = {16},
  year = {2014},
  month = {Feb},
  publisher = {American Physical Society},
  doi = {10.1103/PhysRevA.89.022317},
  url = {https://link.aps.org/doi/10.1103/PhysRevA.89.022317}
}

@article{2018quantumsensor,
  title = {Multiparameter Estimation in Networked Quantum Sensors},
  author = {Proctor, Timothy J. and Knott, Paul A. and Dunningham, Jacob A.},
  journal = {Phys. Rev. Lett.},
  volume = {120},
  issue = {8},
  pages = {080501},
  numpages = {6},
  year = {2018},
  month = {Feb},
  publisher = {American Physical Society},
  doi = {10.1103/PhysRevLett.120.080501},
  url = {https://link.aps.org/doi/10.1103/PhysRevLett.120.080501}
}

@article{2004MultipartyQSS,
  title = {Efficient multiparty quantum-secret-sharing schemes},
  author = {Xiao, Li and Lu Long, Gui and Deng, Fu-Guo and Pan, Jian-Wei},
  journal = {Phys. Rev. A},
  volume = {69},
  issue = {5},
  pages = {052307},
  numpages = {5},
  year = {2004},
  month = {May},
  publisher = {American Physical Society},
  doi = {10.1103/PhysRevA.69.052307},
  url = {https://link.aps.org/doi/10.1103/PhysRevA.69.052307}
}

@article{2003SimonIrvine,
  author  = {Simon, Christoph and Irvine, William T. M.},
  title   = {Robust Long-Distance Entanglement and a Loophole-Free Bell Test with Ions and Photons},
  journal = {Phys. Rev. Lett.},
  volume  = {91},
  issue   = {11},
  pages   = {110405},
  year    = {2003},
  doi     = {10.1103/PhysRevLett.91.110405}
}

@article{Mermin1990,
   author = {Mermin, N. D.},
   title = {Quantum Mysteries Revisited},
   journal = {Am. J. Phys.},
   volume = {58},
   number = {8},
   pages = {731},
   ISSN = {0002-9505},
   DOI = {Doi 10.1119/1.16503},
   url = {<Go to ISI>://WOS:A1990DP88500010},
   year = {1990},
   type = {Journal Article}
}

@article{2024SensingGuha,
  title = {Utilizing probabilistic entanglement between sensors in quantum networks},
  author = {Van Milligen, Emily A. and Gagatsos, Christos N. and Kaur, Eneet and Towsley, Don and Guha, Saikat},
  journal = {Phys. Rev. Appl.},
  volume = {22},
  issue = {6},
  pages = {064085},
  numpages = {22},
  year = {2024},
  month = {Dec},
  publisher = {American Physical Society},
  doi = {10.1103/PhysRevApplied.22.064085},
  url = {https://link.aps.org/doi/10.1103/PhysRevApplied.22.064085}
}

@article{2019Gorshkov,
  title = {Heisenberg-scaling measurement protocol for analytic functions with quantum sensor networks},
  author = {Qian, Kevin and Eldredge, Zachary and Ge, Wenchao and Pagano, Guido and Monroe, Christopher and Porto, J. V. and Gorshkov, Alexey V.},
  journal = {Phys. Rev. A},
  volume = {100},
  issue = {4},
  pages = {042304},
  numpages = {8},
  year = {2019},
  month = {Oct},
  publisher = {American Physical Society},
  doi = {10.1103/PhysRevA.100.042304},
  url = {https://link.aps.org/doi/10.1103/PhysRevA.100.042304}
}

@article{2021Alexey,
  author  = {Qian, Timothy and Bringewatt, Jacob and Boettcher, Igor and Bienias, Przemyslaw and Gorshkov, Alexey V.},
  title   = {Optimal measurement of field properties with quantum sensor networks},
  journal = {Phys. Rev. A},
  volume  = {103},
  issue   = {3},
  pages   = {L030601},
  year    = {2021},
  doi     = {10.1103/PhysRevA.103.L030601}
}

@article{OReilly2024,
  title = {Fast Photon-Mediated Entanglement of Continuously Cooled Trapped Ions for Quantum Networking},
  author = {O'Reilly, Jameson and Toh, George and Goetting, Isabella and Saha, Sagnik and Shalaev, Mikhail and Carter, Allison L. and Risinger, Andrew and Kalakuntla, Ashish and Li, Tingguang and Verma, Ashrit and Monroe, Christopher},
  journal = {Phys. Rev. Lett.},
  volume = {133},
  issue = {9},
  pages = {090802},
  numpages = {6},
  year = {2024},
  month = {Aug},
  publisher = {American Physical Society},
  doi = {10.1103/PhysRevLett.133.090802},
  url = {https://link.aps.org/doi/10.1103/PhysRevLett.133.090802}
}

@article{Saha2025,
  author  = {Saha, S. and Shalaev, M. and O'Reilly, J. and Goetting, I. and Toh, G. and Kalakuntla, A. and Yu, Y. and Monroe, C.},
  title   = {High-fidelity remote entanglement of trapped atoms mediated by time-bin photons},
  journal = {Nat. Commun.},
  volume  = {16},
  pages   = {2533},
  year    = {2025},
  doi     = {10.1038/s41467-025-57557-4}
}

@article{Allcock_2016,
doi = {10.1088/1367-2630/18/2/023043},
url = {https://doi.org/10.1088/1367-2630/18/2/023043},
year = {2016},
month = {feb},
publisher = {IOP Publishing},
volume = {18},
number = {2},
pages = {023043},
author = {Allcock, D T C and Harty, T P and Sepiol, M A and Janacek, H A and Ballance, C J and Steane, A M and Lucas, D M and Stacey, D N},
title = {Dark-resonance Doppler cooling and high fluorescence in trapped Ca-43 ions at intermediate magnetic field},
journal = {New J. Phys.}
}

@article{EIT2016,
  title = {Electromagnetically-induced-transparency ground-state cooling of long ion strings},
  author = {Lechner, Regina and Maier, Christine and Hempel, Cornelius and Jurcevic, Petar and Lanyon, Ben P. and Monz, Thomas and Brownnutt, Michael and Blatt, Rainer and Roos, Christian F.},
  journal = {Phys. Rev. A},
  volume = {93},
  issue = {5},
  pages = {053401},
  numpages = {10},
  year = {2016},
  month = {May},
  publisher = {American Physical Society},
  doi = {10.1103/PhysRevA.93.053401},
  url = {https://link.aps.org/doi/10.1103/PhysRevA.93.053401}
}

@article{cohertime2021,
  author  = {Wang, Pengfei and Luan, Chun-Yang and Qiao, Mu and Um, Mark and Zhang, Junhua and Wang, Ye and Yuan, Xiao and Gu, Mile and Zhang, Jingning and Kim, Kihwan},
  journal = {Nat. Commun.},
  title   = {Single ion qubit with estimated coherence time exceeding one hour},
  volume  = {12},
  pages   = {233},
  year    = {2021},
  doi     = {10.1038/s41467-020-20330-w}
}

@article{Zeemancoherence2016,
  author  = {Ruster, T. and Schmiegelow, C. T. and Kaufmann, H. and Warschburger, C. and Schmidt-Kaler, F. and Poschinger, U. G.},
  title   = {A long-lived {Zeeman} trapped-ion qubit},
  journal = {Appl. Phys. B},
  volume  = {122},
  pages   = {254},
  year    = {2016},
  doi     = {10.1007/s00340-016-6527-4}
}

@article{SPAM_Sotirova2024,
  author = {Sotirova, A. S. and Leppard, J. D. and Vazquez-Brennan, A. and Decoppet, S. M. and Pokorny, F. and Malinowski, M. and Ballance, C. J.},
  title = {High-fidelity heralded quantum state preparation and measurement},
  journal = {arXiv:2409.05805},
  year = {2024},
  url = {https://arxiv.org/abs/2409.05805}
}

@article{highFgates2025,
  title = {Scalable, High-Fidelity All-Electronic Control of Trapped-Ion Qubits},
  author = {L\"oschnauer, C.M. and Mosca Toba, J. and Hughes, A.C. and King, S.A. and Weber, M.A. and Srinivas, R. and Matt, R. and Nourshargh, R. and Allcock, D.T.C. and Ballance, C.J. and Matthiesen, C. and Malinowski, M. and Harty, T.P.},
  journal = {PRX Quantum},
  volume = {6},
  issue = {4},
  pages = {040313},
  numpages = {13},
  year = {2025},
  month = {Oct},
  publisher = {American Physical Society},
  doi = {10.1103/h4wk-v31j},
  url = {https://link.aps.org/doi/10.1103/h4wk-v31j}
}

@article{highF1QG2025,
  title = {Single-Qubit Gates with Errors at the ${10}^{\ensuremath{-}7}$ Level},
  author = {Smith, M. C. and Leu, A. D. and Miyanishi, K. and Gely, M. F. and Lucas, D. M.},
  journal = {Phys. Rev. Lett.},
  volume = {134},
  issue = {23},
  pages = {230601},
  numpages = {8},
  year = {2025},
  month = {Jun},
  publisher = {American Physical Society},
  doi = {10.1103/42w2-6ccy},
  url = {https://link.aps.org/doi/10.1103/42w2-6ccy}
}

@article{Crocker19,
author = {C. Crocker and M. Lichtman and K. Sosnova and A. Carter and S. Scarano and C. Monroe},
journal = {Opt. Express},
number = {20},
pages = {28143},
publisher = {Optica Publishing Group},
title = {High purity single photons entangled with an atomic qubit},
volume = {27},
month = {Sep},
year = {2019},
url = {https://opg.optica.org/oe/abstract.cfm?URI=oe-27-20-28143},
doi = {10.1364/OE.27.028143}
}

@article{Stephenson2020,
  title = {High-Rate, High-Fidelity Entanglement of Qubits Across an Elementary Quantum Network},
  author = {Stephenson, L. J. and Nadlinger, D. P. and Nichol, B. C. and An, S. and Drmota, P. and Ballance, T. G. and Thirumalai, K. and Goodwin, J. F. and Lucas, D. M. and Ballance, C. J.},
  journal = {Phys. Rev. Lett.},
  volume = {124},
  issue = {11},
  pages = {110501},
  numpages = {6},
  year = {2020},
  month = {Mar},
  publisher = {American Physical Society},
  doi = {10.1103/PhysRevLett.124.110501},
  url = {https://link.aps.org/doi/10.1103/PhysRevLett.124.110501}
}

@ARTICLE{Van_Leent_atom_ent_2022,
author={van Leent, Tim
and Bock, Matthias
and Fertig, Florian
and Garthoff, Robert
and Eppelt, Sebastian
and Zhou, Yiru
and Malik, Pooja
and Seubert, Matthias
and Bauer, Tobias
and Rosenfeld, Wenjamin
and Zhang, Wei
and Becher, Christoph
and Weinfurter, Harald},
title={Entangling single atoms over 33{\thinspace}km telecom fibre},
journal={Nature},
year={2022},
month={Jul},
day={01},
volume={607},
number={7917},
pages={69-73},
issn={1476-4687},
doi={10.1038/s41586-022-04764-4},
url={https://doi.org/10.1038/s41586-022-04764-4}
}

@article{Drmota2023,
  title = {Robust Quantum Memory in a Trapped-Ion Quantum Network Node},
  author = {Drmota, P. and Main, D. and Nadlinger, D. P. and Nichol, B. C. and Weber, M. A. and Ainley, E. M. and Agrawal, A. and Srinivas, R. and Araneda, G. and Ballance, C. J. and Lucas, D. M.},
  journal = {Phys. Rev. Lett.},
  volume = {130},
  issue = {9},
  pages = {090803},
  numpages = {7},
  year = {2023},
  month = {Mar},
  publisher = {American Physical Society},
  doi = {10.1103/PhysRevLett.130.090803},
  url = {https://link.aps.org/doi/10.1103/PhysRevLett.130.090803}
}

@article{Bernien2013,
   author = {Bernien, H. and Hensen, B. and Pfaff, W. and Koolstra, G. and Blok, M. S. and Robledo, L. and Taminiau, T. H. and Markham, M. and Twitchen, D. J. and Childress, L. and Hanson, R.},
   title = {Heralded entanglement between solid-state qubits separated by three metres},
   journal = {Nature},
   volume = {497},
   number = {7447},
   pages = {86–90},
   ISSN = {0028-0836},
   DOI = {10.1038/nature12016},
   url = {<Go to ISI>://WOS:000318221500038
https://www.nature.com/articles/nature12016.pdf},
   year = {2013},
   type = {Journal Article}
}

@article{Kumar2023,
doi = {10.1088/1367-2630/acdd1a},
url = {https://doi.org/10.1088/1367-2630/acdd1a},
year = {2023},
month = {jun},
publisher = {IOP Publishing},
volume = {25},
number = {6},
pages = {063027},
author = {Kumar, Shreya and Bhatti, Daniel and Jones, Alex E and Barz, Stefanie},
title = {Experimental entanglement generation using multiport beam splitters},
journal = {New Journal of Physics}
}

@article{tritter,
  author  = {Spagnolo, N. and Vitelli, C. and Aparo, L. and Mataloni, P. and Sciarrino, F. and Crespi, A. and Ramponi, R. and Osellame, R.},
  title   = {Three-photon bosonic coalescence in an integrated tritter},
  journal = {Nat. Commun.},
  volume  = {4},
  pages   = {1606},
  year    = {2013},
  doi     = {10.1038/ncomms2616},
  type = {Journal Article},
     url = {<Go to ISI>://WOS:000318873900060
https://www.nature.com/articles/ncomms2616.pdf}
}

@article{quantumrepeater1998,
  title = {Quantum Repeaters: The Role of Imperfect Local Operations in Quantum Communication},
  author = {Briegel, H.-J. and D\"ur, W. and Cirac, J. I. and Zoller, P.},
  journal = {Phys. Rev. Lett.},
  volume = {81},
  issue = {26},
  pages = {5932--5935},
  numpages = {0},
  year = {1998},
  month = {Dec},
  publisher = {American Physical Society},
  doi = {10.1103/PhysRevLett.81.5932},
  url = {https://link.aps.org/doi/10.1103/PhysRevLett.81.5932}
}

@article{sympathetic,
  author  = {Sakrejda, Tomasz P. and Zhukas, Liudmila A. and Blinov, Boris B.},
  title   = {Efficient sympathetic cooling in mixed barium and ytterbium ion chains},
  journal = {Quantum Inf. Process.},
  volume  = {20},
  pages   = {162},
  year    = {2021},
  doi     = {10.1007/s11128-021-03112-1}
}

@article{fairsampling,
  title = {Hidden-Variable Example Based upon Data Rejection},
  author = {Pearle, Philip M.},
  journal = {Phys. Rev. D},
  volume = {2},
  issue = {8},
  pages = {1418--1425},
  numpages = {0},
  year = {1970},
  month = {Oct},
  publisher = {American Physical Society},
  doi = {10.1103/PhysRevD.2.1418},
  url = {https://link.aps.org/doi/10.1103/PhysRevD.2.1418}
}

@article{MerminIneq1990,
  title = {Extreme quantum entanglement in a superposition of macroscopically distinct states},
  author = {Mermin, N. David},
  journal = {Phys. Rev. Lett.},
  volume = {65},
  issue = {15},
  pages = {1838--1840},
  numpages = {0},
  year = {1990},
  month = {Oct},
  publisher = {American Physical Society},
  doi = {10.1103/PhysRevLett.65.1838},
  url = {https://link.aps.org/doi/10.1103/PhysRevLett.65.1838}
}

@article{Rowe2001,
author={Rowe, M. A.
and Kielpinski, D.
and Meyer, V.
and Sackett, C. A.
and Itano, W. M.
and Monroe, C.
and Wineland, D. J.},
title={Experimental violation of a Bell's inequality with efficient detection},
journal={Nature},
year={2001},
month={Feb},
day={01},
volume={409},
number={6822},
pages={791-794},
doi={10.1038/35057215},
url={https://doi.org/10.1038/35057215}
}

@article{Sackett2000,
   author = {Sackett, C. A. and Kielpinski, D. and King, B. E. and Langer, C. and Meyer, V. and Myatt, C. J. and Rowe, M. and Turchette, Q. A. and Itano, W. M. and Wineland, D. J. and Monroe, C.},
   title = {Experimental entanglement of four particles},
   journal = {Nature},
   volume = {404},
   number = {6775},
   pages = {256},
   ISSN = {0028-0836},
   DOI = {Doi 10.1038/35005011},
   url = {<Go to ISI>://WOS:000086022200039
https://www.nature.com/articles/35005011.pdf},
   year = {2000},
   type = {Journal Article}
}

@Article{Moehring2007,
journal={Nature},
author={D. L. Moehring and P. Maunz and S. Olmschenk and K. C. Younge and D. N. Matsukevich and L.-M. Duan and C. Monroe},
title={Entanglement of single-atom quantum bits at a distance},
year={2007},
month={September},
pages={68-71},
volume={449},
number={7158},
doi={10.1038/nature06118},
url={https://ideas.repec.org/a/nat/nature/v449y2007i7158d10.1038_nature06118.html},
}

@article{Nolleke2013,
  title = {Efficient Teleportation Between Remote Single-Atom Quantum Memories},
  author = {N\"olleke, Christian and Neuzner, Andreas and Reiserer, Andreas and Hahn, Carolin and Rempe, Gerhard and Ritter, Stephan},
  journal = {Phys. Rev. Lett.},
  volume = {110},
  issue = {14},
  pages = {140403},
  numpages = {5},
  year = {2013},
  month = {Apr},
  publisher = {American Physical Society},
  doi = {10.1103/PhysRevLett.110.140403},
  url = {https://link.aps.org/doi/10.1103/PhysRevLett.110.140403}
}

@Article{Delteil2016,
author={Delteil, Aymeric
and Sun, Zhe
and Gao, Wei-bo
and Togan, Emre
and Faelt, Stefan
and Imamo{\u{g}}lu, Ata{\c{c}}},
title={Generation of heralded entanglement between distant hole spins},
journal={Nature Phys.},
year={2016},
month={Mar},
day={01},
volume={12},
number={3},
pages={218-223},
issn={1745-2481},
doi={10.1038/nphys3605},
url={https://doi.org/10.1038/nphys3605}
}

@article{Awschalom2021,
  title = {Development of Quantum Interconnects (QuICs) for Next-Generation Information Technologies},
  author = {Awschalom, David and others},
  journal = {PRX Quantum},
  volume = {2},
  issue = {1},
  pages = {017002},
  numpages = {21},
  year = {2021},
  month = {Feb},
  publisher = {American Physical Society},
  doi = {10.1103/PRXQuantum.2.017002},
  url = {https://link.aps.org/doi/10.1103/PRXQuantum.2.017002}
}

@article{HOM,
  title = {Measurement of subpicosecond time intervals between two photons by interference},
  author = {Hong, C. K. and Ou, Z. Y. and Mandel, L.},
  journal = {Phys. Rev. Lett.},
  volume = {59},
  issue = {18},
  pages = {2044--2046},
  numpages = {0},
  year = {1987},
  month = {Nov},
  publisher = {American Physical Society},
  doi = {10.1103/PhysRevLett.59.2044},
  url = {https://link.aps.org/doi/10.1103/PhysRevLett.59.2044}
}

@article{Yu2026,
  title = {Entanglement-fidelity limits of photonically networked atomic qubits from recoil and timing},
  author = {Yu, Yichao and Saha, Sagnik and Shalaev, Mikhail and Toh, George and O'Reilly, Jameson and Goetting, Isabella and Kalakuntla, Ashish and Monroe, Christopher},
  journal = {Phys. Rev. A},
  volume = {113},
  issue = {1},
  pages = {012620},
  numpages = {9},
  year = {2026},
  month = {Jan},
  publisher = {American Physical Society},
  doi = {10.1103/82c6-4nkl},
  url = {https://link.aps.org/doi/10.1103/82c6-4nkl}
}

@article{confkey,
  author  = {Pickston, A. and Ho, J. and Ulibarrena, A. and Grasselli, F. and Proietti, M. and Morrison, C. L. and Barrow, P. and Graffitti, F. and Fedrizzi, A.},
  title   = {Conference key agreement in a quantum network},
  journal = {npj Quantum Inf.},
  volume  = {9},
  pages   = {82},
  year    = {2023},
  doi     = {10.1038/s41534-023-00750-4}
}

@article{threepartyRNG,
  title = {Device-independent quantum secret sharing with advanced random key generation basis},
  author = {Zhang, Qi and others},
  journal = {Phys. Rev. A},
  volume = {111},
  issue = {1},
  pages = {012603},
  numpages = {14},
  year = {2025},
  month = {Jan},
  publisher = {American Physical Society},
  doi = {10.1103/PhysRevA.111.012603},
  url = {https://link.aps.org/doi/10.1103/PhysRevA.111.012603}
}

\clearpage

\section*{Appendices}\label{suppinfo}

\subsection{Phase tracking}\label{phasetrackingsupp}
Let's first consider one qubit, say in node A. After excitation and successful collection into fiber, but before the beam splitter (BS), the state becomes:
\begin{equation}
    \ket{\psi} = \frac{1}{\sqrt{2}} (e^{i(k_Hx_H - (\omega_H + \omega_{\downarrow} )t)} \ket{\downarrow H} + e^{i(k_Vx_V -(\omega_V + \omega_{\uparrow }) t)} \ket{\uparrow V})
\end{equation}
\begin{equation}
    \ket{\psi} = \frac{1}{\sqrt{2}} e^{-i \omega t} (e^{ik_Hx_H} \ket{\downarrow H} + e^{ik_Vx_V} \ket{\uparrow V})
\end{equation}
where $\downarrow/\uparrow$ refers to the internal ion state and $H/V$ refers to the polarization of the single photon. Here, $k$ refers to the wavenumber of emission, $x$ is the distance the photon travels from the ion to the beamsplitter, $\omega_{V/H}$ is the frequency of the excited to ground state transition, and $\omega_{\downarrow/\uparrow}$ is the frequency of the ion's energy level. The qubit frequency is $\omega = \omega_H + \omega_{\downarrow} = \omega_V + \omega_{\uparrow}$. The phase from the qubit frequency $\omega$ is global, so we can factor it out and ignore it in this discussion. We can rewrite this as:
\begin{equation*}
    \ket{\psi} = \frac{1}{\sqrt{2}} (\ket{\downarrow H} + e^{i (k_V x_V - k_H x_H)} \ket{\uparrow V})
\end{equation*}
\begin{equation}
    \ket{\psi} = \frac{1}{\sqrt{2}} (\ket{\downarrow H} + e^{i \Delta k x} \ket{\uparrow V})
\end{equation}
Here $\Delta k = k_V - k_H$ is the difference in the emission wave numbers between the $H$ and $V$ photons.
The path length $x$ is the same for $H/V$: $x = x_V = x_H$. 

Now, let's consider the entire system in which we collect a photon into fiber from each chamber:
\begin{equation*}
    \ket{\Psi} = \ket{\psi}_{A} \otimes \ket{\psi}_{B} \otimes \ket{\psi}_{C}
\end{equation*}

\begin{multline} \label{eq:ABCionphoton}
    \ket{\Psi} = \frac{1}{2\sqrt{2}} (\ket{\downarrow_A H_A} + e^{i \Delta k_Ax_A} \ket{\uparrow _A V_A}) \\ \otimes (\ket{\downarrow_B H_B} + e^{i\Delta k_B x_B} \ket{\uparrow _B V_B}) \\\otimes (\ket{\downarrow_C H_C} + e^{i \Delta k_C x_C} \ket{\uparrow _C V_C})
\end{multline}
where $I = A,B,C$ denotes the ion as well as the corresponding PBS input port of the emitted photon, as referenced in Figure~\ref{fig:GHZsetup}.

We represent the half-wave plate transformation as:
\begin{equation*}
    \ket{H}_l \rightarrow \frac{1}{\sqrt{2}} (\ket{H}_{l'} + \ket{V}_{l'})
\end{equation*}
\begin{equation*}
    \ket{V}_l \rightarrow \frac{1}{\sqrt{2}} (\ket{H}_{l'} - \ket{V}_{l'})
\end{equation*}
Here, $l'$ corresponds to the output ports of the beam splitters.

Starting from Eqn. \ref{eq:ABCionphoton}, after going through the PBS, applying the half-wave plate transformations, ignoring normalization, and before detection:
\begin{gather*}
    \ket{\Psi} \rightarrow  [(\ket{H}_3 + \ket{V}_3) \ket{\downarrow}_A + e^{i \Delta k_Ax_A}(-\ket{H}_1 + \ket{V}_1) \ket{\uparrow }_A] \\
    \otimes [(\ket{H}_1 + \ket{V}_1) \ket{\downarrow}_B + e^{i \Delta k_Bx_B}(\ket{H}_2 - \ket{V}_2) \ket{\uparrow }_B] \\
     \otimes [(\ket{H}_2 + \ket{V}_2) \ket{\downarrow}_C + e^{i \Delta k_Cx_C}(-\ket{H}_3 + \ket{V}_3) \ket{\uparrow }_C]
\end{gather*}
Note that the change in sign for the $\ket{V}$ transformation comes from the PBS reflection. 

Once a photon from a specific chamber is detected, there will be a relative phase from the two energy levels of the ion qubit precessing differently from each other. We introduce this as $e^{-i \omega_{I,i} t_{d}}$ where $I$ corresponds to an ion \{A,B,C\}, $i$ refers to the internal state of the respective ion, $\omega$ is the ion qubit frequency, and $t_{d}$ is the time when a detector $d$ clicks. 
\begin{multline*}
\ket{\Psi} \rightarrow
\Bigl[(e^{-i\omega_{A,0}t_{3H}}\ket{H}_{3H} - e^{-i\omega_{A,0}t_{3V}}\ket{V}_{3V})\ket{\downarrow}_A \\
+ e^{i\Delta k_A x_A}(-e^{-i\omega_{A,1}t_{\uparrow H}}\ket{H}_{\uparrow H} - e^{-i\omega_{A,1}t_{\uparrow V}}\ket{V}_{\uparrow V})\ket{\uparrow }_A\Bigr] \\
\otimes\Bigl[(e^{-i\omega_{B,0}t_{\uparrow H}}\ket{H}_{\uparrow H} - e^{-i\omega_{B,0}t_{\uparrow V}}\ket{V}_{\uparrow V})\ket{\downarrow}_B \\
+ e^{i\Delta k_B x_B}(e^{-i\omega_{B,1}t_{2H}}\ket{H}_{2H} + e^{-i\omega_{B,1}t_{2V}}\ket{V}_{2V})\ket{\uparrow }_B\Bigr] \\
\otimes\Bigl[(e^{-i\omega_{C,0}t_{2H}}\ket{H}_{2H} - e^{-i\omega_{C,0}t_{2V}}\ket{V}_{2V})\ket{\downarrow}_C \\
+ e^{i\Delta k_C x_C}(-e^{-i\omega_{C,1}t_{3H}}\ket{H}_{3H} - e^{-i\omega_{C,1}t_{3V}}\ket{V}_{3V})\ket{\uparrow }_C\Bigr]
\end{multline*}
Consider click pattern (1H, 2H, 3H), which heralds the plus GHZ state:
\begin{multline*} 
    \ket{\Psi} \rightarrow  [(e^{-i \omega_{A,0} t_{3H}}\ket{H}_{3H}) \ket{\downarrow}_A \\
    + e^{i \Delta k_Ax_A} (-e^{-i \omega_{A,1} t_{\uparrow H}}\ket{H}_{\uparrow H}) \ket{\uparrow }_A] \\
    \otimes [(e^{-i \omega_{B,0} t_{\uparrow H}} \ket{H}_{\uparrow H} ) \ket{\downarrow}_B \\+ e^{i \Delta k_Bx_B}(e^{-i \omega_{B,1} t_{2H}}\ket{H}_{2H}) \ket{\uparrow }_B] \\
     \otimes [(e^{-i \omega_{C,0} t_{2H}}\ket{H}_{2H})\ket{\downarrow}_C \\+ e^{i \Delta k_Cx_C} (-e^{-i \omega_{C,1} t_{3H}}\ket{H}_{3H})
     \ket{\uparrow }_C]
\end{multline*}

Ignoring double detections (when two photons hit the same detector at the same time),
\begin{gather*} 
    \rightarrow  (e^{-i \omega_{A,0} t_{3H}}\ket{H}_{3H}) \ket{\downarrow}_A \\
    \otimes(e^{-i \omega_{B,0} t_{\uparrow H}} \ket{H}_{\uparrow H} ) \ket{\downarrow}_B\\
    \otimes(e^{-i \omega_{C,0} t_{2H}}\ket{H}_{2H})\ket{\downarrow}_C \\ +
    e^{i \Delta k_Ax_A} (-e^{-i \omega_{A,1} t_{\uparrow H}}\ket{H}_{\uparrow H}) \ket{\uparrow }_A \\
    \otimes e^{i \Delta k_Bx_B}(e^{-i \omega_{B,1} t_{2H}}\ket{H}_{2H}) \ket{\uparrow }_B \\
    \otimes e^{i \Delta k_Cx_C} (-e^{-i \omega_{C,1} t_{3H}}\ket{H}_{3H})
     \ket{\uparrow }_C
     \end{gather*}
     
\begin{gather*} 
    \rightarrow  \ket{H}_{3H} \ket{\downarrow}_A 
     \ket{H}_{\uparrow H} \ket{\downarrow}_B
    \ket{H}_{2H} \ket{\downarrow}_C +
    \\
    e^{i (\Delta k_Ax_A + \Delta k_Bx_B + \Delta k_Cx_C)} \\
    \otimes (e^{-i (\omega_{A,1} - \omega_{B,0} )t_{\uparrow H}}\ket{H}_{\uparrow H}) \ket{\uparrow }_A \\
    \otimes (e^{-i (\omega_{B,1} - \omega_{C,0}) t_{2H}}\ket{H}_{2H}) \ket{\uparrow }_B \\
    \otimes (e^{-i (\omega_{C,1} - \omega_{A,0}) t_{3H}}\ket{H}_{3H} \ket{\uparrow }_C
     \end{gather*}

Assuming perfect magnetic field matching, 
\begin{gather*} 
    \rightarrow   [\ket{\downarrow\downarrow \downarrow}
    \\ + 
    e^{(\Delta kx_A + \Delta kx_B + \Delta kx_C) } e^{-i \omega (t_{\uparrow H} + t_{2H} + t_{3H})} \ket{\uparrow \uparrow \uparrow}]
    \ket{HHH}
     \end{gather*}
where the total phase is \begin{equation*}
    \Phi = \Delta k(x_A+x_B+x_C) -  \omega (t_{1H} + t_{2H} + t_{3H})
\end{equation*}

The path lengths $x_A,x_B,x_C$ change by a maximum of $<100 ~\mu$m between shots, and with a $\Delta k\approx0.4$ rad/m, the phase variation between shots from the $\Delta kx$ terms is negligible. The phase from the $\omega t$ terms is much more significant, 
as $\omega\approx 2\pi(11.9)$ MHz, leading to appreciable phase accrual with times $t$ on the nanosecond scale ($14$ ns gives $\approx1$ rad of phase). This necessitates proper phase tracking at the level of $\leq 1$ ns to ensure that the phase of the entangled state between successive experimental shots remains consistent. Our current phase tracking of 1 ns precision gives an infidelity of $< 0.005$. 

The phase correction proceeds as follows. First, we record the times $t_{1H},t_{2H},t_{3H}$ at which the detectors click. We then calculate the average of these times $t_{\text{avg}}=(t_{1H}+t_{2H}+t_{3H})/3$ and delay our analysis pulse sequence by this $t_{\text{avg}}$. By delaying the global analysis pulse sequence by $t_{\text{avg}}$, each of the qubits experience a phase evolution of $\Delta\Phi_{1H/2H/3H} = w(t_{\text{const}}+t_{\text{avg}}-t_{1H/2H/3H})$, where $t_{\text{const}}=3.5~\mu$s is the time between the start of an entanglement attempt and the start of our analysis pulse sequence given the attempt is a success. Adding the $\Delta\Phi$ accrued by each of the qubits gives a total phase accrued on our entangled state of $\Delta\Phi = w(3t_{\text{const}}+3t_{\text{avg}}-t_{1H}-t_{2H}-t_{3H})=3wt_{\text{const}}$. As  $t_{\text{const}}$ is consistent from shot to shot, our phase tracking ensures that the entangled phase does not vary significantly between experiment shots. 

\subsection{GHZ-state fidelity bounds}\label{fidboundsupp}

\noindent The GHZ fidelity for a general mixed state is:
\begin{gather}
    \mathcal{F} = \bra{\text{GHZ}}\rho\ket{\text{GHZ}} \notag \\= \frac{1}{2}\left(\rho_{\downarrow\downarrow\downarrow,\downarrow\downarrow\downarrow} + \rho_{\uparrow\uparrow\uparrow,\uparrow\uparrow\uparrow} + 2\,\text{Re}[e^{-i\Phi}\rho_{\downarrow\downarrow\downarrow,\uparrow\uparrow\uparrow}]\right),
\end{gather}
where $\Phi$ is the phase of the target GHZ state. The populations $\rho_{\downarrow\downarrow\downarrow,\downarrow\downarrow\downarrow}$ and $\rho_{\uparrow\uparrow\uparrow,\uparrow\uparrow\uparrow}$ are directly measurable; it remains to extract $\rho_{\downarrow\downarrow\downarrow,\uparrow\uparrow\uparrow}$ from a parity scan.

\medskip
\noindent\textbf{General phase rotations.} We apply $R(\phi_1,\phi_2,\phi_3) \equiv R(\pi/2, \phi_1)\otimes R(\pi/2, \phi_2)\otimes R(\pi/2, \phi_3)$ to each qubit before measuring $Z\otimes Z\otimes Z$, where
\begin{equation}
    R(\theta, \phi) = 
    \begin{pmatrix}
        \cos\tfrac{\theta}{2} & -ie^{-i\phi}\sin\tfrac{\theta}{2}\\
        -ie^{i\phi}\sin\tfrac{\theta}{2} & \cos\tfrac{\theta}{2}
    \end{pmatrix}
    \label{eq:general_rotation}
\end{equation}
is a general single-qubit rotation. The parity expectation value is:
\begin{gather}
    \langle \hat{\Pi}(\phi_1,\phi_2,\phi_3) \rangle = \mathrm{Tr}\![\rho \cdot R^\dagger(\phi_1,\phi_2,\phi_3)\left(Z\otimes Z\otimes Z\right) \notag \\R(\phi_1,\phi_2,\phi_3)].
\end{gather}

\noindent The single-qubit matrix elements are: %$R^\dagger(\pi/2, \phi_k)ZR(\pi/2, \phi_k)$ are:
\begin{align}
    \bra{\downarrow}R^\dagger(\pi/2, \phi_k)ZR(\pi/2,\phi_k)\ket{\downarrow} &=  0, \notag\\
    \bra{\uparrow}R^\dagger(\pi/2, \phi_k)ZR(\pi/2, \phi_k)\ket{\uparrow} & = 0 \notag \\
    \bra{\downarrow}R^\dagger(\pi/2, \phi_k)ZR(\pi/2,\phi_k)\ket{\uparrow} &= -ie^{-i\phi_k}, \notag \\
    \bra{\uparrow}R^\dagger(\pi/2, \phi_k)ZR(\pi/2,\phi_k)\ket{\downarrow} &= ie^{i\phi_k}. \notag
\end{align}
Since diagonal single-qubit elements vanish, only weight-3 coherences (where all three qubit indices differ) survive. The full expectation value is:
\begin{align}
    \langle \hat{\Pi}(\phi_1,\phi_2,\phi_3) \rangle =\;&
    -ie^{i(\phi_1+\phi_2+\phi_3)}\rho_{\downarrow\downarrow\downarrow,\uparrow\uparrow\uparrow}
    \notag\\&+ ie^{-i(\phi_1+\phi_2+\phi_3)}\rho_{\uparrow\uparrow\uparrow,\downarrow\downarrow\downarrow} \notag\\
    &+ ie^{i(-\phi_1+\phi_2+\phi_3)}\rho_{\uparrow\downarrow\downarrow,\downarrow\uparrow\uparrow}
    \notag\\&- ie^{-i(-\phi_1+\phi_2+\phi_3)}\rho_{\downarrow\uparrow\uparrow,\uparrow\downarrow\downarrow} \notag\\
    &+ ie^{i(\phi_1-\phi_2+\phi_3)}\rho_{\downarrow\uparrow\downarrow,\uparrow\downarrow\uparrow}
    \notag\\&- ie^{-i(\phi_1-\phi_2+\phi_3)}\rho_{\uparrow\downarrow\uparrow,\downarrow\uparrow\downarrow} \notag\\
    &+ ie^{i(\phi_1+\phi_2-\phi_3)}\rho_{\downarrow\downarrow\uparrow,\uparrow\uparrow\downarrow}
    \notag\\&- ie^{-i(\phi_1+\phi_2-\phi_3)}\rho_{\uparrow\uparrow\downarrow,\downarrow\downarrow\uparrow}. \notag
\end{align}

\noindent Writing $\rho_{abc,a'b'c'} = |\rho_{abc,a'b'c'}|e^{i\Phi_{abc,a'b'c'}}$ and collecting conjugate pairs:
\begin{align}
    &\notag \langle \hat{\Pi}(\phi_1,\phi_2,\phi_3) \rangle = \\ &+2|\rho_{\downarrow\downarrow\downarrow,\uparrow\uparrow\uparrow}|\sin(\phi_1+\phi_2+\phi_3 + \Phi_{\downarrow\downarrow\downarrow,\uparrow\uparrow\uparrow}) \notag\\
    &+2|\rho_{\uparrow\downarrow\downarrow,\downarrow\uparrow\uparrow}|\sin(\phi_1-\phi_2-\phi_3 + \Phi_{\uparrow\downarrow\downarrow,\downarrow\uparrow\uparrow}) \notag\\
    &+2|\rho_{\downarrow\uparrow\downarrow,\uparrow\downarrow\uparrow}|\sin(-\phi_1+\phi_2-\phi_3 + \Phi_{\downarrow\uparrow\downarrow,\uparrow\downarrow\uparrow}) \notag\\
    &+2|\rho_{\downarrow\downarrow\uparrow,\uparrow\uparrow\downarrow}|\sin(-\phi_1-\phi_2+\phi_3 + \Phi_{\downarrow\downarrow\uparrow,\uparrow\uparrow\downarrow}).
\end{align}
\noindent Each coherence oscillates at a different linear combination of the three phases.

\noindent\textbf{Three-phase scan.} Setting $\phi_1 = \phi_2 = \phi_3 = \phi$, the four terms oscillate at frequencies $3\phi$, $\phi$, $\phi$, and $\phi$ respectively. Isolating the $3\phi$ Fourier component we find:
\begin{equation}
    \langle \hat{\Pi}(\phi,\phi,\phi) \rangle\big|_{3\phi} = 2|\rho_{\downarrow\downarrow\downarrow,\uparrow\uparrow\uparrow}|\sin(3\phi + \Phi_{\downarrow\downarrow\downarrow,\uparrow\uparrow\uparrow}),
\end{equation}
with amplitude  $|\rho_{\downarrow\downarrow\downarrow,\uparrow\uparrow\uparrow}|$ and phase $\Phi_{\downarrow\downarrow\downarrow,\uparrow\uparrow\uparrow}$, so $\text{Re}[\rho_{\downarrow\downarrow\downarrow,\uparrow\uparrow\uparrow}] = |\rho_{\downarrow\downarrow\downarrow,\uparrow\uparrow\uparrow}|\cos(\Phi_{\downarrow\downarrow\downarrow,\uparrow\uparrow\uparrow})$. Choosing the target state phase $\Phi \equiv \Phi_{\downarrow\downarrow\downarrow,\uparrow\uparrow\uparrow}$,
\begin{equation}
\boxed{
    \mathcal{F} = \frac{1}{2}\left(\rho_{\downarrow\downarrow\downarrow,\downarrow\downarrow\downarrow} + \rho_{\uparrow\uparrow\uparrow,\uparrow\uparrow\uparrow} + 2|\rho_{\downarrow\downarrow\downarrow,\uparrow\uparrow\uparrow}|\right).
}
\end{equation}

\medskip
\noindent\textbf{Single-qubit scan and fidelity bounds.} Setting instead $\phi_1 = \phi$ and $\phi_2 = \phi_3 = 0$, all four coherence terms oscillate at $\phi$:
\begin{align}
    \langle \hat{\Pi}(\phi,0,0) \rangle =\;&
    +2|\rho_{\downarrow\downarrow\downarrow,\uparrow\uparrow\uparrow}|\sin(\phi + \Phi_{\downarrow\downarrow\downarrow,\uparrow\uparrow\uparrow})
    \notag\\&+2|\rho_{\uparrow\downarrow\downarrow,\downarrow\uparrow\uparrow}|\sin(\phi + \Phi_{\uparrow\downarrow\downarrow,\downarrow\uparrow\uparrow}) \notag\\
    &-2|\rho_{\downarrow\uparrow\downarrow,\uparrow\downarrow\uparrow}|\sin(\phi - \Phi_{\downarrow\uparrow\downarrow,\uparrow\downarrow\uparrow})
    \notag\\&-2|\rho_{\downarrow\downarrow\uparrow,\uparrow\uparrow\downarrow}|\sin(\phi - \Phi_{\downarrow\downarrow\uparrow,\uparrow\uparrow\downarrow}).
\end{align}

\noindent The measured oscillation amplitude, or contrast $\mathcal{C}$, now satisfies
\begin{equation}
    \mathcal{C} \leq 2|\rho_{\downarrow\downarrow\downarrow,\uparrow\uparrow\uparrow}| + 2|\rho_{\uparrow\downarrow\downarrow,\downarrow\uparrow\uparrow}| + 2|\rho_{\downarrow\uparrow\downarrow,\uparrow\downarrow\uparrow}| + 2|\rho_{\downarrow\downarrow\uparrow,\uparrow\uparrow\downarrow}|.
\end{equation}

\noindent Applying the Cauchy--Schwarz inequality to the non-target coherences, $|\rho_{abc,a'b'c'}| \leq \sqrt{\rho_{abc,abc}\,\rho_{a'b'c',a'b'c'}}$, we obtain a lower bound on $|\rho_{\downarrow\downarrow\downarrow,\uparrow\uparrow\uparrow}|$ from the single-qubit scan:
\begin{equation}
\boxed{
\begin{aligned}
    &\mathcal{F} \geq \frac{1}{2}\!\Bigl[\rho_{\downarrow\downarrow\downarrow,\downarrow\downarrow\downarrow} + \rho_{\uparrow\uparrow\uparrow,\uparrow\uparrow\uparrow} + \mathcal{C} \\
    &- 2\Bigl(\sqrt{\rho_{\uparrow\downarrow\downarrow,\uparrow\downarrow\downarrow}\,\rho_{\downarrow\uparrow\uparrow,\downarrow\uparrow\uparrow}}
    + \sqrt{\rho_{\downarrow\uparrow\downarrow,\downarrow\uparrow\downarrow}\,\rho_{\uparrow\downarrow\uparrow,\uparrow\downarrow\uparrow}} \\
    &+ \sqrt{\rho_{\downarrow\downarrow\uparrow,\downarrow\downarrow\uparrow}\,\rho_{\uparrow\uparrow\downarrow,\uparrow\uparrow\downarrow}} \Bigr)\Bigr]
\end{aligned}
}
\end{equation}

\noindent The upper bound arises when the contrast $\mathcal{C}$ satisfies
\begin{equation}
    \mathcal{C} \geq 2|\rho_{\downarrow\downarrow\downarrow,\uparrow\uparrow\uparrow}| - 2|\rho_{\uparrow\downarrow\downarrow,\downarrow\uparrow\uparrow}| - 2|\rho_{\downarrow\uparrow\downarrow,\uparrow\downarrow\uparrow}| - 2|\rho_{\downarrow\downarrow\uparrow,\uparrow\uparrow\downarrow}|.
\end{equation}
By applying the Cauchy-Schwarz inequality as above, we obtain an upper bound on $|\rho_{\downarrow\downarrow\downarrow,\uparrow\uparrow\uparrow}|$ from the single-qubit scan:
\begin{equation}
\boxed{
\begin{aligned}
    \mathcal{F} \leq \frac{1}{2}\Bigl[ &\rho_{\downarrow\downarrow\downarrow,\downarrow\downarrow\downarrow} + \rho_{\uparrow\uparrow\uparrow,\uparrow\uparrow\uparrow} + \mathcal{C} \\
    &+ 2\Bigl( \sqrt{\rho_{\uparrow\downarrow\downarrow,\uparrow\downarrow\downarrow}\,\rho_{\downarrow\uparrow\uparrow,\downarrow\uparrow\uparrow}}
    + \sqrt{\rho_{\downarrow\uparrow\downarrow,\downarrow\uparrow\downarrow}\,\rho_{\uparrow\downarrow\uparrow,\uparrow\downarrow\uparrow}} \\
    &+ \sqrt{\rho_{\downarrow\downarrow\uparrow,\downarrow\downarrow\uparrow}\,\rho_{\uparrow\uparrow\downarrow,\uparrow\uparrow\downarrow}} \Bigr) \Bigr]
\end{aligned}
}
\end{equation}

We measure populations of 0.955(8) and an average parity of 0.771(21), giving a bounded fidelity of $0.841(17) \leq \mathcal{F} \leq 0.881(17)$.

\subsection{Parity characterization}\label{paritysupp}
\begin{figure}[h!]
  \centering
\includegraphics[width=0.5\textwidth]{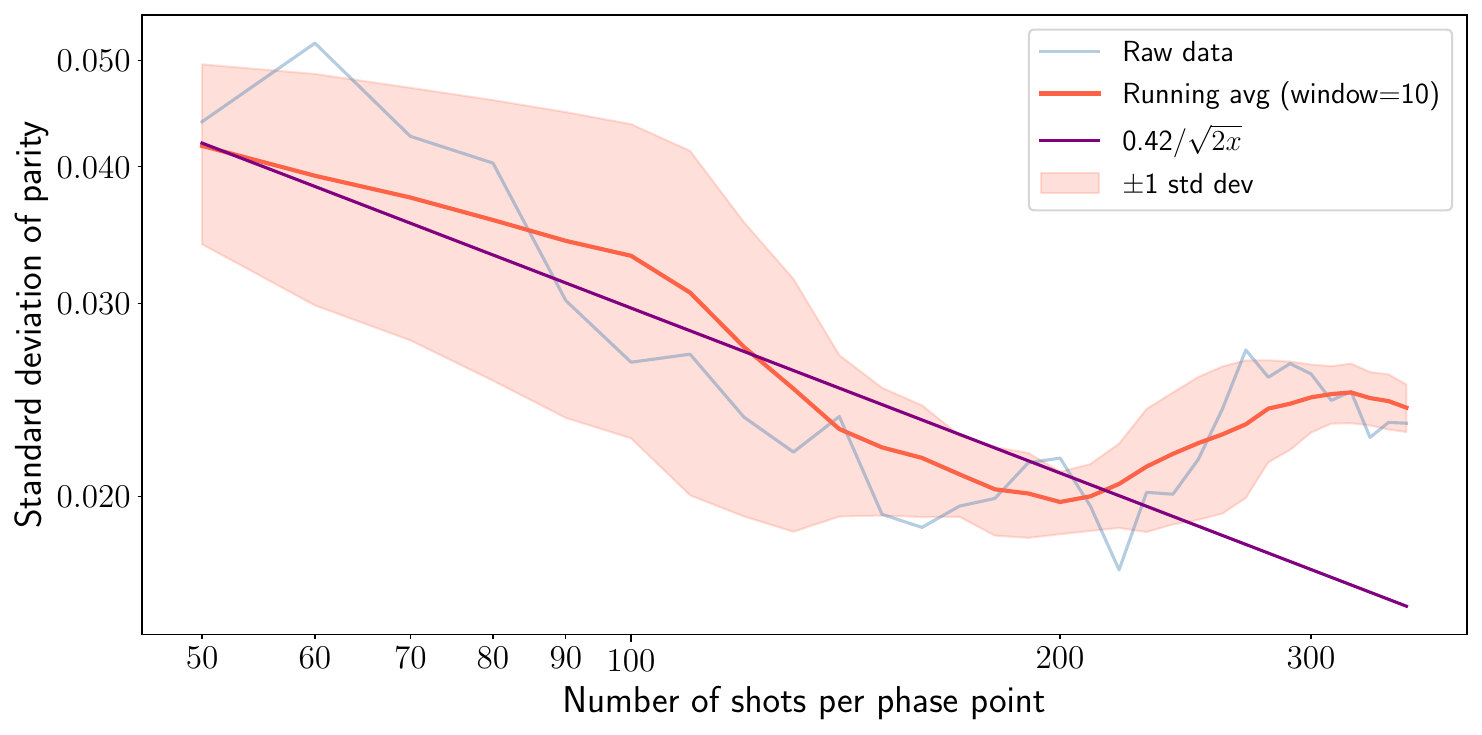}
  \caption{\textbf{Standard deviation of concatenated parity} Standard deviation of parity over concatenated data points of all runs. The running average is binned in windows of 10, and the pre-factor for the $1/\sqrt{2x}$ line comes from the variance of the binomial distribution of the total parity value P=0.771. Pre-factor $=\sqrt{P(1-P)} = 0.42$.}
  \label{Fig:stddev}
\end{figure}

\begin{figure*}[t]
\phantomsection \centering \includegraphics[width=0.75\textwidth]{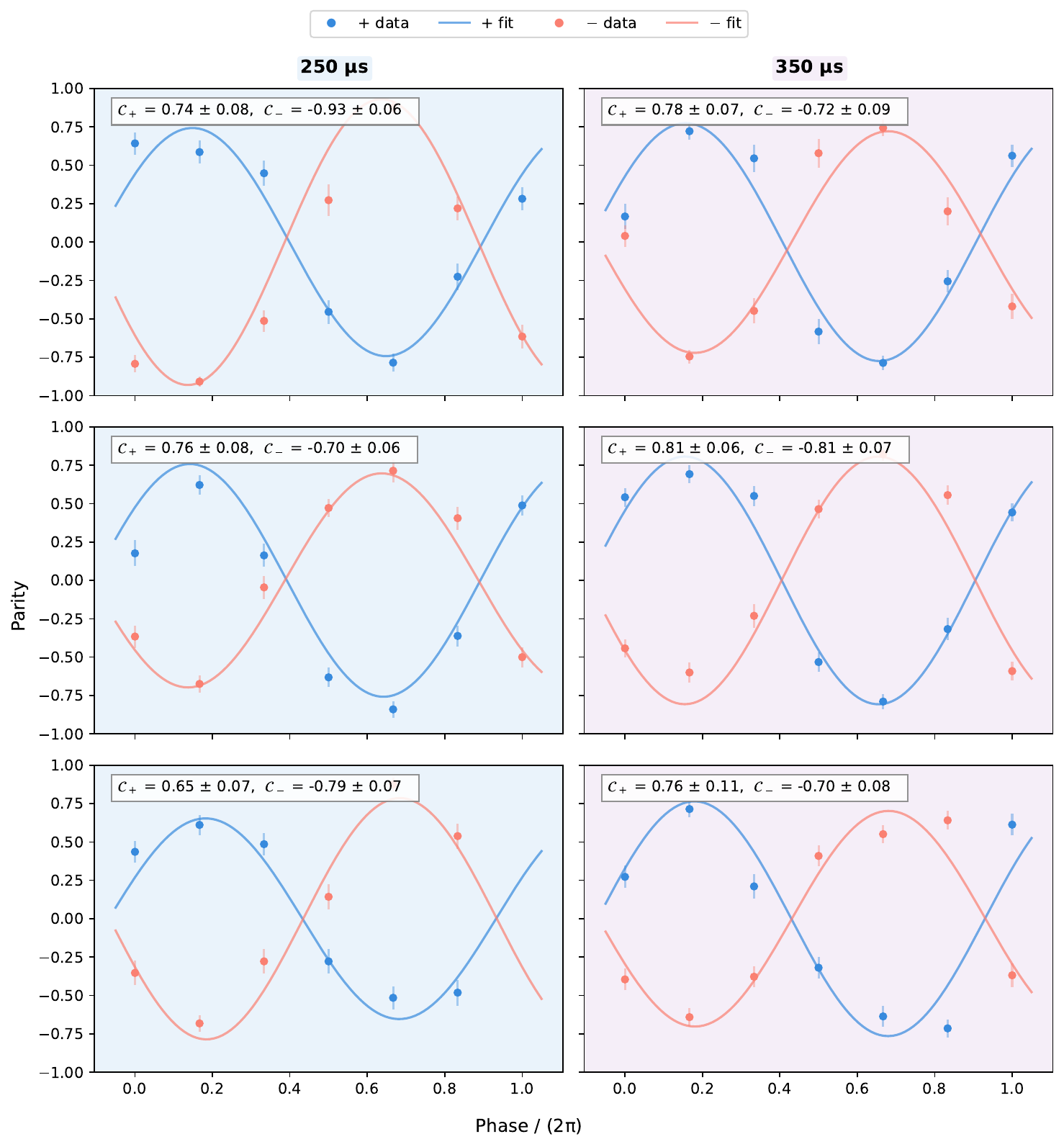}
  \caption{\textbf{Parity datasets} All six parity datasets taken for the results in this paper. The left columns are for 250 $\mu$s fast loop time and the right columns are for 350 $\mu$s fast loop time.}
  \label{Fig:parity}
\end{figure*}

Figure~\ref{Fig:parity} shows all six data runs taken across two days for two different entangling attempt times. The total average parity reported in the main text is obtained by averaging the fits of the individual runs. 
Figure~\ref{Fig:stddev} justifies this approach: when the data from all runs are concatenated, the uncertainty on the fitted parity initially decreases as $1/\sqrt{2x}$, consistent with shot-noise-limited statistics. It reaches a minimum at approximately 200 data points before increasing again, indicating that the runs are not statistically identical. We therefore average the fits of the individual runs rather than fitting the concatenated data.

\subsection{Photon indistinguishability}\label{indistinguishabilitysupp}

Temporal overlap is matched to $<100$~ps by using a Time-Correlated Single Photon Counting (TCSPC) device to look at the arrival times of pairs of photons per detector. The 493~nm pulsed light fiber-couplers are on translation stages such that adjusting the distance the pulsed light travels changes the arrival timing accordingly. Frequency overlap between the photons is ensured by calibrating the magnetic field before each run and between parity points.

The polarization degree-of-freedom defines the photonic qubits. The first sets of QWP/HWP on the GHZ-generator are calibrated to maximize ion-photon correlations for each input. This effectively undoes the unique unitary imposed on each photon via their respective fiber, rotating the H and V polarizations to the intended H/V basis. The measured ion-photon correlations for $A/B/C$ are 0.989(2), 0.990(1), and 0.963(2), respectively.

We now show how imperfect spatial mode-matching of the photons affects the maximum GHZ fidelity. 
A monochromatic photon with general polarization ($\sigma$) and Laguerre--Gauss mode $(p = 0, \ell = 0)$ is written as
\begin{equation}
    \hat{a}^\dagger_{\sigma}|0\rangle = \int d\vec{\rho}\; LG_{0,0}(\rho,z)\,\hat{a}^\dagger_\sigma(\hat{r})|0\rangle,
    \qquad \vec{r} = \rho\,\hat{\rho} + z\hat{z}.
\end{equation}
We assume each detector is insensitive to frequency and spatial mode, but sensitive to a single polarization, as the detectors are always placed after a polarizing beam splitter. The action of the $k$-th detector is then
\begin{equation}
    \hat{\Pi}_{\sigma_{k}} = \int d\phi\,d\rho \rho\; \hat{a}^\dagger_{\sigma_k}(\vec{\rho}\,,z_k)|0\rangle\langle 0|\,\hat{a}_{\sigma_k}(\vec{\rho}\,,z_k),
\end{equation}
where the integral is over the detection plane, $z_k$ is the detector position and $\sigma_{k}$ the measured polarization. As shown in Fig. \ref{fig:GHZsetup}, a successful GHZ measurement requires a 3-photon coincidence. Suppose we detect clicks on detectors $d_{1}, \  d_{2} $ and $d_{3}$. The heralded three ion state is then
\begin{equation}
    \rho_{\text{ion}} = \frac{\text{Tr}_{\text{photon}} \left[ \hat{\Pi}_{\sigma_{d_{1}}}\hat{\Pi}_{\sigma_{d_{2}}}\hat{\Pi}_{\sigma_{d_{3}}} \rho_{\text{ion-photon}} \right]}{\text{Tr} \left[ \hat{\Pi}_{\sigma_{d_{1}}}\hat{\Pi}_{\sigma_{d_{2}}}\hat{\Pi}_{\sigma_{d_{3}}} \rho_{\text{ion-photon}} \right]}
\end{equation}

The GHZ state fidelity then becomes
\begin{equation}
    \mathcal{F} = \frac{1}{2} + \frac{V^{BA}_{\sigma_{d_{1}}}    V^{CB}_{\sigma_{d_{2}}} V^{AC}_{\sigma_{d_{3}}}}{V^{BB}_{\sigma_{d_{1}}}V^{CC}_{\sigma_{d_{2}}}V^{AA}_{\sigma_{d_{3}}} + V^{AA}_{\sigma_{d_{1}}}V^{BB}_{\sigma_{d_{2}}}V^{CC}_{\sigma_{d_{3}}}}
\end{equation}
with the overlap integrals
\begin{equation}
\begin{aligned}
    V_k^{ij} & \equiv  \int \rho d\rho d\phi \int \rho'd\rho'd\phi'\int\rho''d\rho'' d\phi''  LG^{*}_{00, i}(\rho', z') \\ & LG_{00, j}(\rho'', z'')  \langle0| a_{\sigma_{k}}(\rho', z')a_{\sigma_{k}}^{\dagger}(\rho, z_{k})|0\rangle \\&\langle0| a_{\sigma_{k}}(\rho, z_{k})a_{\sigma_{k}}^{\dagger}(\rho'', z'')|0\rangle,
\end{aligned}
\end{equation}
where the indices $i, j = A, B, C$ denote the photon-mode and $\sigma_{k}$ the detected polarization. 

For simplicity we assume all photon modes and detectors share the same $\hat{z}$. To align the coordinate systems we shift
\begin{equation}
    \vec{\rho}' \mapsto \vec{\rho}^{'} + \vec{\delta}_{ij}/2, \qquad \vec{\rho}^{''} \mapsto \vec{\rho}^{''} - \vec{\delta}_{ij}/2,
\end{equation}
where $\vec{\delta}_{ij}$ is the transverse separation between two photon mode axes at the detector plane. This introduces delta functions $\delta(z'-z_k)\,\delta(z''-z_k)\,\delta(\vec{\rho}-\vec{\rho}''-\vec{\delta}_{ij}/2)$, so
\begin{equation}
    V_k^{ij} =\int \rho d\rho\,d\phi\; LG_{00, i}^*(\vec{\rho}-\vec{\delta}_{ij}/2,z_k)\,LG_{00, j}(\vec{\rho}+\vec{\delta}_{ij}/2,z_k),
\end{equation}
which is the overlap integral of two Gaussians whose centres are separated by $\vec{\delta}_{ij}$, evaluated at the detector plane $z_k$. With this simplification, the GHZ fidelity becomes
\begin{equation}
    \mathcal{F} = \frac{1}{2} + \frac{V^{BA}\,V^{CB}\,V^{AC}}{2},
\end{equation}
where $V^{ij}$ denotes the normalized Gaussian overlap between beams $i$ and $j$ at the detector plane.

We measure the visibilities as follows. Continuous-wave 493~nm laser light is coupled into two out of the three fibers for pairwise alignment. This light is first split into two paths that each travel through an in-fiber AOM with a 10~MHz beat note between them, effectively creating an interferometer. Each path is then coupled into a fiber and the resultant fringe pattern at each output arm gives an estimate of the pairwise visibility. A beam profiler is used to view and adjust the spatial overlap at a close (tens of centimeters) and large (a few meters) distance from the single-photon detectors. 
We measure pairwise visibilities $\bar{V}$ of $0.9724/0.9838/0.9838$ for each output arm, respectively. These measurements are limited by laser coherence, so they provide an upper bound for the error from spatial mode mismatch. 

\subsection{Ion-ion entanglement on the GHZ-state generator}\label{ionionsupp}

As a way to debug and characterize the system, ion-ion entanglement between nodes A and B was done on the GHZ-state generator (inputs 1 and 2). Not pictured in Fig. \ref{fig:GHZsetup} is a HWP in-between the two middle PBS's. Nominally the HWP is set to $0^{\circ}$ during GHZ experiments such that it does not impart any rotations on the single photons. However, to do ion-ion entanglement, it is set to $22.5^{\circ}$ such that this waveplate is the one that erases the ``which-path information" instead of the HWPs at output modes 2 and 3, which are now set to $0^{\circ}$. The output HWP at mode 1 is unchanged at $22.5^{\circ}$. This simple modification allows ion-ion entanglement between the first two inputs. 

\begin{figure}[h!]
\phantomsection
  \centering
  \includegraphics[width=0.5\textwidth]{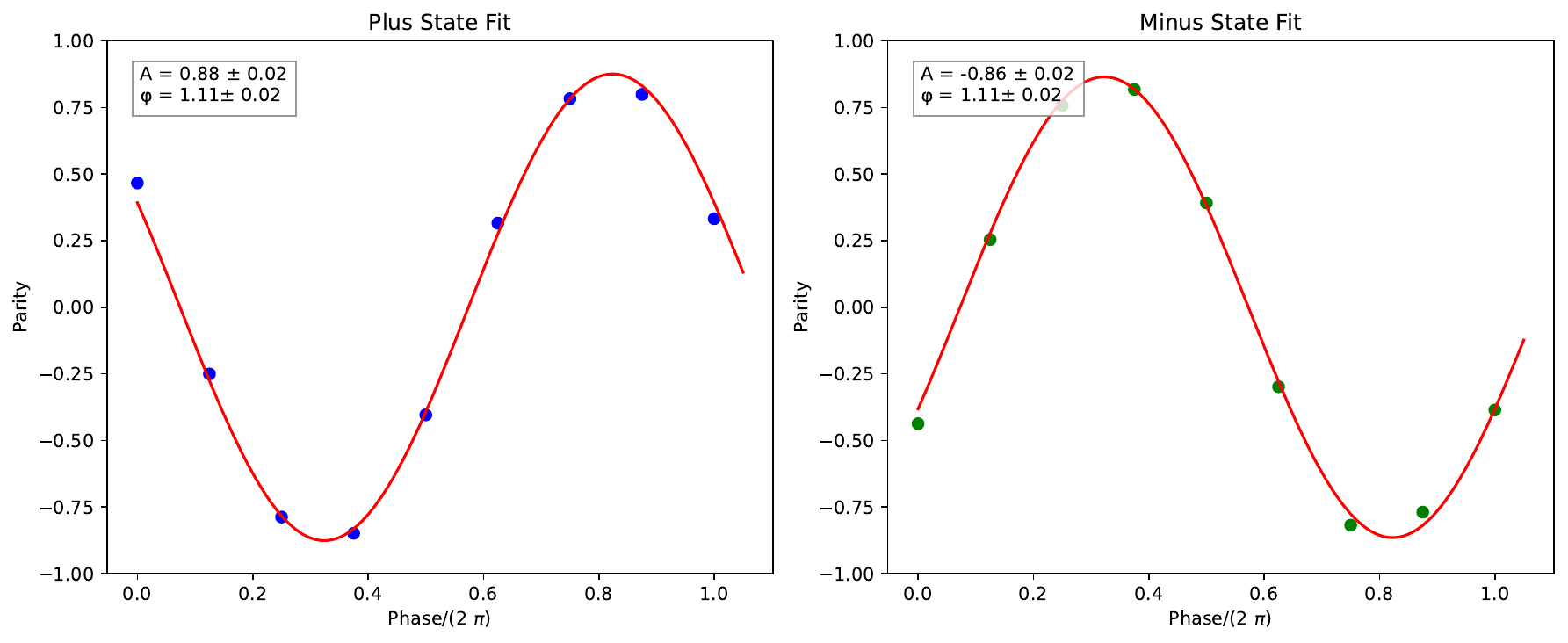}
  \includegraphics[width=0.5\textwidth]{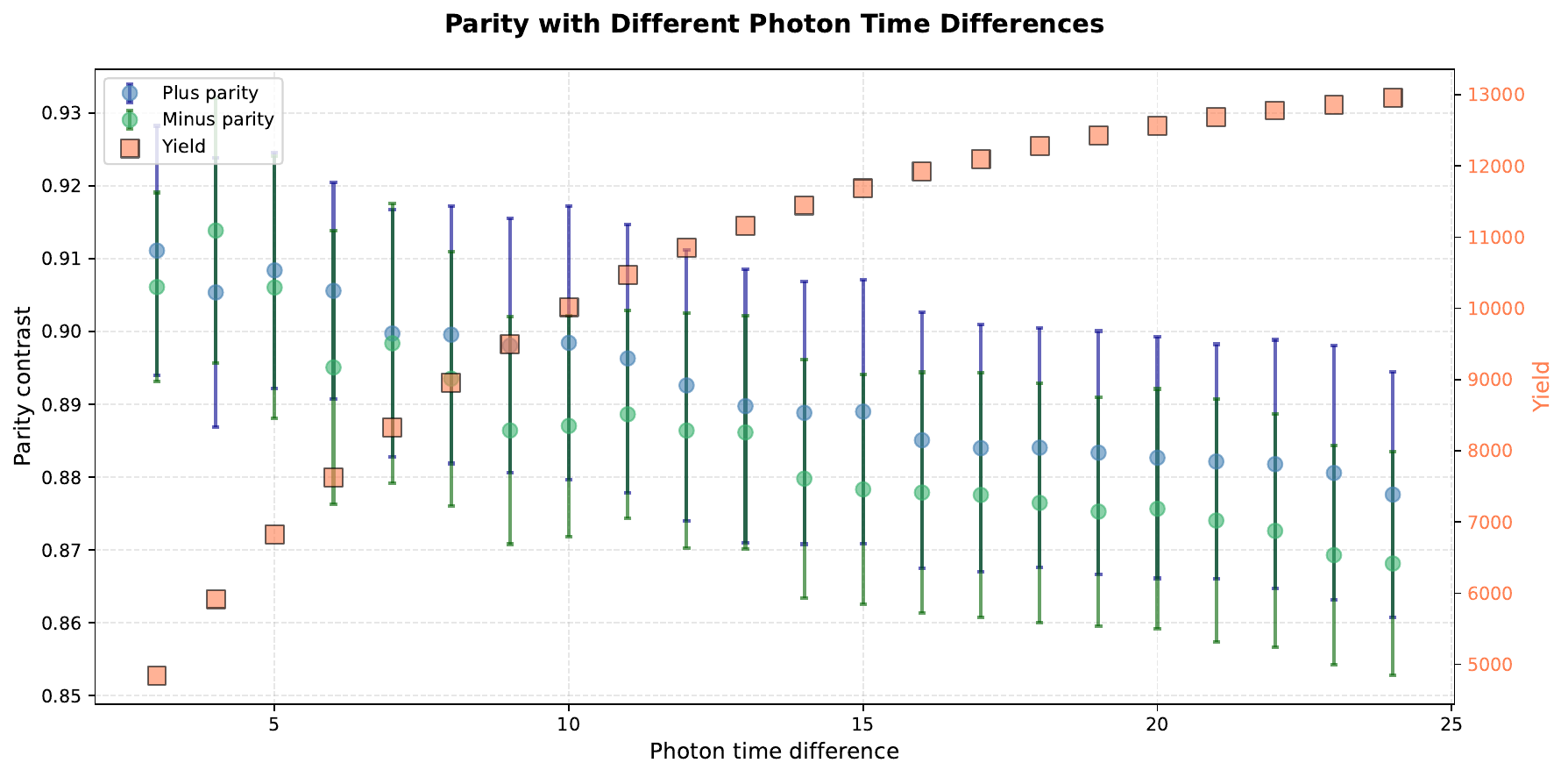}
  \caption{\textbf{Ion-ion parity} Error bars are statistical and are too small to be seen on this plot. 13,368 data points were taken over a clock time of about 15 minutes.  \\ An increase in parity contrast is observed when the parity results of ion-ion entanglement between A/B are filtered by the difference in arrival times of the photons. This is due to averaging over the different phases of the residual entangled spin-motion ion states. For this data set, there is a $4(1) \%$ difference of parity in total.}
  \label{fig:ionion}
\end{figure}

The ion-ion entanglement experimental sequence is identical to the GHZ sequence except for the heralding patterns. We measure populations of 0.982(2) and an average parity of 0.870(11), giving a total fidelity of 0.926(6). The average success probability was $28.1(2) \times 10^{-6}$ and the average entanglement generation rate was $18.4(2) \sec^{-1}$. Due to averaging over many phases from the ions being in different thermal states during the 50~ns detection window, the resulting entangled state fidelity decreases \cite{Yu2026}. Figure \ref{fig:ionion} shows this effect for ion-ion entanglement between nodes $A$ and $B$, which results in about a $0.04(1)$ parity decrease. Because node $C$ has higher systematic infidelities and worse cooling, as shown in its success probability, this result is a lower bound on the total error contribution to the GHZ-state fidelity. 

\subsection{Error budget}\label{errorbudgetsupp}
Given the measured pairwise visibilities of 0.9724/0.9838/0.9838, the GHZ-state infidelity is bounded to a maximum of $0.03(1)$. 
Ion-photon coherence error is mainly due to polarization mixing, with a small contribution from 1762~nm SPAM error. The measured correlations/coherences for A/B/C are 0.989(2)/0.985(1), 0.990(1)/0.987(1), and 0.963(2)/0.965(1), respectively. The main error contributing to correlation infidelity is polarization mixing, and the difference between the fidelity of correlations and coherences is error from 1762~nm SPAM. These two effects contribute an infidelity of $0.052(2)$. 

Other error sources that contribute less than one percent to the infidelity of the GHZ state include temporal overlap of the photons, different magnetic fields at each ion, and waveplate calibration on the GHZ-state generator. 
There is about $11 \mu$s after heralding a successful photon event and the analysis $\pi/2$ pulse, and the phase tracking has a precision of 1~ns. Also, with a coherence time of $350 \mu$s, decoherence during this idle time is negligible. Temporal overlap of the photons are matched to better than $50-100$ ps, resulting in a maximum error of $\exp[-(\Delta t_{AB} + \Delta t_{AC} +\Delta t_{BC})/2 t_{\text{decay}}] = 0.006$. Wave plates are calibrated to a precision of $\epsilon =1^{\circ}$, and the error goes as $\cos^4{(\epsilon)}$ (three output HWPs and the middle HWP discussed in Appendix \ref{ionionsupp}), giving an error bound of $< 0.005$. The probability of obtaining a double-excitation is $ {e^{-0.49*.003/7.85}}$, or 0.0002. The APDs have an average dark count rate of 10 Hz, so the probability that one pair of the three pairs of detectors gets a dark count in the 50~ns photon window is $< 10^{-6}$. Thus, these error sources are rendered negligible on the total state fidelity. 

\subsection{Rate of GHZ state generation}\label{GHZratesupp}

The measured average success probability is $0.159(3) \times10^{-6}$ and the measured average entanglement rate is $0.093(5)\sec^{-1}$, giving an average duty cycle of $D \approx 59\% $. The discrepancy between our measured GHZ success probability of $0.159(3) \times10^{-6}$ and our expected success probability of $\frac{1}{4}RDp_A p_B p_C=0.252\times10^{-6}$ is largely due to fiber coupling drift between calibrations and ion decrystallizing during long runs. We also note that fiber coupling in C is lower than previous works due to high ion temperature. Further cooling investigations have increased fiber coupling by a factor of $1.3$ to the nominal $30 \%$. Table \ref{tab:rate} shows this effect; for longer entanglement attempt time, the duty cycle increases but the success probability decreases from heating effects, ultimately resulting in a similar rate for the two different attempt times. The average of the two duty cycles is about $66 \%$, but the measured average duty cycle is $59 \%$, indicating some level of decrystallization occurrence. 

\begin{table}[h!]
\begin{tabular}{|c|c|c|c|}
\hline
\textbf{\begin{tabular}[c]{@{}c@{}}Entanglement \\ attempt time \\ ($\mu$s)\end{tabular}} & \multicolumn{1}{l|}{\textbf{\begin{tabular}[c]{@{}l@{}}Duty \\ cycle\end{tabular}}}  & \textbf{\begin{tabular}[c]{@{}c@{}}$ \frac{1}{4} p_{A} p_B p_C$\\ $\times 10^{-6}$\end{tabular}} & \textbf{\begin{tabular}[c]{@{}c@{}}$r_{ent}$\\ $\sec^{-1}$\end{tabular}} \\ \hline\hline
250 & 0.625 & 0.180(5)& 0.095(8)  \\ \hline
350 & 0.70 & 0.147(4)& 0.091(7)  \\ \hline
\end{tabular}
\caption{Measured success probability and entanglement rate for different entanglement attempt times. }
\label{tab:rate}
\end{table}

\begin{figure}[h!]
  \centering   
    \includegraphics[width=0.5\textwidth]{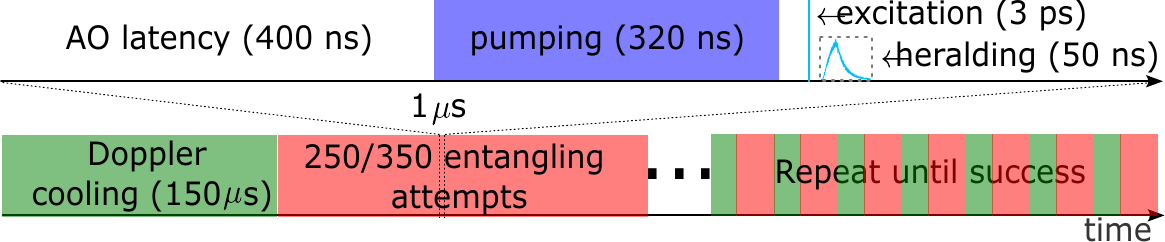}
  \caption{\textbf{Experimental Sequence} Each $1 \mu$s entangling attempt consists of 400 ns of AOM latency, 320 ns of optical pumping, fast $493$~nm excitation, and the 50 ns photon heralding window. This process is repeated 250/350 times or until a successful event is recorded, followed by $150 \mu$s of Doppler cooling as described in Methods. 
  }
  \label{fig:expseq}
\end{figure}

\end{document}